\documentstyle[12pt,aaspp4,flushrt]{article}

\newcommand{\be}{\begin{equation}}
\newcommand{\ee}{\end{equation}}

\begin{document}

\title{SINGULAR ISOTHERMAL DISKS: \\
II. NONAXISYMMETRIC BIFURCATIONS AND EQUILIBRIA}

\author{Daniele Galli}
\affil{Osservatorio Astrofisico di Arcetri \\
              Largo Enrico Fermi 5 \\
              I-50125 Firenze, Italy \\
              galli@arcetri.astro.it}

\author{Frank H. Shu}
\affil{Department of Astronomy, University of California \\
              Berkeley, CA 94720, USA \\
              shu@vshu.berkeley.edu}

\author{Gregory Laughlin}
\affil{NASA/Ames Research Center \\
              MS 245-3, Moffett Field, CA 94035\\
              gpl@acetylene.arc.nasa.gov}

\author{Susana Lizano}
\affil{Instituto de Astronom\'{\i}a, UNAM \\
              Apdo 70-264 \\
              4510 M\'exico, D.F., Mexico \\
              lizano@astrosmo.unam.mx}

\begin{abstract}

We review the difficulties of the classical fission and fragmentation
hypotheses for the formation of binary and multiple stars.  A crucial
missing ingredient in previous theoretical studies is the inclusion of
dynamically important levels of magnetic fields.  As a minimal model
for a candidate presursor to the formation of binary and multiple
stars, we therefore formulate and solve the problem of the equilibria
of isopedically magnetized, singular isothermal disks, without the
assumption of axial symmetry.  Considerable analytical progress can be
made if we restrict our attention to models that are scale-free, i.e.,
that have surface densities that vary inversely with distance $\varpi$
from the rotation axis of the system.  In agreement with earlier
analysis by Syer and Tremaine, we find that lopsided ($M=1$)
configurations exist at any dimensionless rotation rate, including
zero.  Multiple-lobed ($M$ = 2, 3, 4, ...) configurations bifurcate
from an underlying axisymmetric sequence at progressively higher
dimensionless rates of rotation, but such nonaxisymmetric sequences
always terminate in shockwaves before they have a chance to fission
into $M=2$, 3, 4, ... separate bodies.  On the basis of our experience
in this paper, and the preceding Paper I, we advance the hypothesis
that binary and multiple star-formation from smooth (i.e., not highly
turbulent) starting states that are supercritical but in unstable
mechanical balance requires the rapid (i.e., dynamical) loss of
magnetic flux at some stage of the ensuing gravitational collapse.

\end{abstract}


\section{Introduction: Figures of Equilibrium and Binary Star Formation}
\label{sec_intr}

\subsection{The Fission Hypothesis}
\label{sub_fiss}

The fission hypothesis for binary star formation evolved from Newton's
calculation in the seventeenth century for the shape of a rotating
Earth.  Newton imagined an ingenious experiment boring holes to the
center of our planet and filling them with water to show that the Earth
is flatter at the poles than at the equator.  This conclusion embroiled
him in controversy with Cassini, who claimed on the basis of
astronomical measurements that the Earth is prolate rather than
oblate.  (See Todhunter 1873 for a more detailed description, in
particular, for an account of Maupertuis's expedition to Lapland that
settled the debate empirically in favor of Newton.)

Newton computed the gravitational field of a spheroid
of small but not negligible eccentricity, with the centrifugal
effects taken into account in the fluid equilibrium.  The general
analytic expression describing the self-consistent eccentricity
$e\equiv \sqrt{1-\ell_3^2/\ell_1^2}$ of an equilibrium spheroid of
uniform density $\rho$ with principal axes $\ell_3 \le \ell_2 = \ell_1$
that rotates with constant angular velocity $\Omega$ was given by
Maclaurin in 1742 
\be
\beta \equiv {\Omega^{2} \over {\pi G \rho}} =
{2{(1-e^{2})}^{1/2} \over {e^{3}}}
(3 -2e^{2}) \, {{\sin}^{-1} e} - {6 \over {e^{2}}}(1-e^{2}) \, .
\ee

In the following year, Simpson (more widely known in connection with
his ``rule'') noticed that the Maclaurin spheroids can exist only if
the rotational parameter $\beta \le 0.449331$ \footnote{As pointed out
by the referee, the correct control parameter for describing a system
of constant mass $M$ and angular momentum $J$, and increasing $\rho$ is
neither $\beta$ nor $\Omega$. The dimensionless control parameter most
like $\beta$ but made from the invariants $J$, $M$ and $G$ and the
density $\rho$ is $(J/M)^2(\rho/M)^{4/3}(1/\pi G\rho)$ which always
increases as $\rho$ increases, whereas $\beta$ has a maximum.}. For
$\beta$ less than this critical value, {\it two} solutions exist, one
more flattened than the other.  At $\beta=0$, these two solutions
correspond to a sphere ($e=0$, most easily imagined in the limit
$\Omega \rightarrow 0$ with $\rho$ finite) and a razor thin disk
($e=1$, most easily imagined in the limit $\rho \rightarrow \infty$
with surface density $\Sigma \equiv \int \rho \, dz$ and $\Omega$
finite).

Ninety one years later, Jacobi (1834) became intrigued by the existence
of two entirely separate equilibria at low $\beta$.  He was
particularly impressed by the fact that the less-obvious disk-like
solution cannot be accessed from the spheroidal solution by means of a
linear perturbation analysis. The presence of two unrelated solutions
suggested to him that others may also exist. Jacobi relaxed the
requirement of axisymmetry and showed that uniformly rotating,
self-gravitating, liquid, masses can also assume triaxial equilibrium
figures in which the principal axes $\ell_{1}$, $\ell_{2}$, and
$\ell_{3}$ have unequal values.

Meyer (in 1842) discovered that the Jacobian sequence of triaxial
ellipsoids branches from the Maclaurin spheroids when the latter's
eccentricity reaches $e=0.81267$ ($\beta=0.37423$). At that point, the
figure axes $\ell_{1}$ and $\ell_{2}$ of the Jacobian ellipsoids become
equal, and Jacobian sequence merges into the Maclaurin sequence.  If a
Maclaurin spheroid is allowed to dissipate energy and contract
homologously to higher density while conserving angular momentum, it
will become triaxial when $e$ exceeds 0.81267.  In other words, the
Maclaurin spheroids are secularly unstable with respect to viscous
forces and bifurcation into Jacobian ellipsoids\footnote{Consult
Chandrasekhar~(1969) for an account of the dynamical instability of
Maclaurin spheroids against transformation into Riemann ellipsoids that
contain internal circulation. He also analyzed the secular instability
of rotating ellipsoids against transformation by gravitational
radiation into Dedekind ellipsoids whose figure axes remain fixed in
space.}.

In 1885, Poincar\'e found that the Jacobian sequence bifurcates into
further classes of equilibrium that have lop-sided shapes.  The first
bifurcation sequence corresponds to a series of egg-shaped figures that
become pear-shaped, and occurs when $\beta=0.28403$.  Poincar\'e
envisioned the slow evolution of a contracting spheroid in which the
contraction time scale is much longer than the internal viscous
timescale so that uniform rotation can be maintained.  Such an object
was imagined to progress along the Maclaurin sequence as it spins up.
Upon reaching $\beta=0.37423$, it would lose its axial symmetry and
become a Jacobian ellipsoid.  Poincar\'e then conjectured that further
secular evolution to $\beta=0.28403$ and beyond would lead to
bifurcation into the pear-shaped sequence of figures, which, in the
face of additional increases in the density and rotation rate, would
eventually fission into a parent body and a satellite, such as the
Earth and its Moon. The same sequence of events was invoked by G.H.
Darwin (1906), the son of the naturalist, to account for the origin of
binary stars (see also Darwin's 1909 review).

Liapounoff~(1905) and Cartan~(1928), however, discovered that the
Jacobi sequence of ellipsoids becomes dynamically unstable at exactly
the point ($\beta=0.28403$) where Poincar\'e's pear-shaped figures
first appear. The inevitable appearance of dynamical instabilities
renders the fission hypothesis problematical, in part because of the
mathematical difficulties associated with describing three-dimensional
nonlinear hydrodynamical evolution.  A more fundamental difficulty
arises from uniformly rotating gaseous equilibrium configurations with
realistic degrees of central condensation (for example, gaseous
polytropes) reaching equatorial breakup prior to bifurcation into
triaxial configurations (James 1964).  Furthermore, if, as likely,
internal viscous timescales exceed the contraction timescale, a
polytropic configuration will develop differential rotation.  As
clarified by Ostriker \& Mark (1968), and Ostriker \& Bodenheimer
(1973), contracting differentially-rotating polytropes become
bar-unstable before reaching equatorial breakup.  Therefore, a
realistic modern descendant of the fission hypothesis would amount to
the conjecture that an unstable barred figure fragments into two or
more pieces. This hypothesis foundered when definitive numerical
simulations by Durisen et al. (1986) demonstrated that the emergent bar
drives spiral waves that transport angular momentum outward and mass
inward, in the process stabilizing the configuration against fission.
Astronomically, this result is consistent with the observation that
bars in flattened galaxies drive outer spiral structures, and do not
spin off additional galaxies.

\subsection{The Fragmentation Hypothesis}
\label{sub_frag} 

An alternative theory for the formation of binary stars can be traced
back to Jeans (1902), who specified the minimum mass, $M_J \propto
G^{-3/2} a^3 \rho^{-1/2}$ for an object of isothermal sound speed $a$
and mean density $\rho$, to collapse under its self-gravity in the
presence of opposing gradients of gas pressure (see also Ebert 1955,
and Bonnor 1955) .  Hoyle (1953) considered a large cloud with mass
$M\sim M_J$ initially.  As it collapses, with $a$ held constant
(because radiative losses under optically thin conditions tend to keep
cosmic gases isothermal) but $\rho$ increasing, the cloud progressively
contains additional Jeans-mass subunits, which might collapse
individually onto their own centers of attraction.  Adjacent collapsing
subfragments could then conceivably wind up as binary stars. A
stability analysis by Hunter (1962) of homogeneously collapsing,
pressure-free spheres seemed to support the Hoyle conjecture.  However,
Layzer (1964) argued that because the overall collapse and the growth
of perturbations proceed with the same powers of time, individual
subunits may have insufficient time to condense into independent
entities before the entire cloud disappeared into the singularity of
Hunter's background state (the analog of the big crunch in a
closed-universe cosmology).

A further difficulty with the fragmentation hypothesis arises because
self-gravitating systems that are initially close to hydrostatic
equilibrium (or have only one Jeans mass) are necessarily centrally
condensed. Numerical calculations by Larson (1969) indicated that such
centrally condensed masses would collapse highly non-homologously.  In
the case of a singular isothermal sphere -- which has a density
distribution $\rho = a^2/2\pi G r^2$ and which contains one Jeans mass
at each radius $r$ -- Shu (1977) showed that collapse proceeds in a
self-similar manner, from ``inside-out''.  Past the moment $t=0$ when
collapse is initiated, a rarefaction wave moves outward at the speed of
sound $a$ into the hydrostatic envelope of the cloud.  At any given
time $t > 0$, roughly half of the disturbed material is infalling, and
half has been incorporated into a tiny hydrostatic central protostar
approximated as a mass point.  At no time in the process does any
subvolume excluding the center contain more than one Jeans mass.  Shu
(1977) conjectured that such solutions are unlikely to fragment, a
conclusion verified by Tohline (1982) to apply more generally to a wide
variety of centrally-condensed collapses.

If such a collapsing cloud is imbued with angular momentum, a structure
containing a star/disk/infalling-envelope naturally develops (Terebey,
Shu \& Cassen 1984).  Numerical work by Boss (1993) removing the
assumption of axial symmetry indicates that rotating collapse flows
with radial density profiles as centrally concentrated as $\rho \propto
r^{-2}$ also avoid fragmentation on the way down.  The fragmentation
hypothesis is therefore restricted either to cases of the collapse of
less centrally condensed clouds (e.g. Burkert, Bate \& Bodenheimer
1997), or else to cases of breakup into multiple gravitating bodies
after a disk has already formed.

Although the issues of gravitational instabilities and fragmentation
within disks are still active areas of investigation, calculations by
Laughlin \& Bodenheimer (1994), which specifically followed the
nonaxisymmetric evolution of disks arising from the collapse of
rotating $r^{-2}$ clouds, did {\it not} find disk fragmentation (see
also Tomley et al. 1994; Pickett, et al. 1998).  Rather, as the disks
arising from the collapse flow become gravitationally unstable, they
develop spiral structures which elicit an inward flux of mass and an
outward flux of angular momentum that proves sufficiently efficient as
to stabilize the disk against fragmentation (see also Laughlin,
Korchagin \& Adams 1998).

Boss (1993) has conjectured that isolated molecular cloud
cores with density laws as steep as $\rho \propto r^{-2}$ will
inevitably lead to the formation of single stars accompanied by planets
rather than binary stars.  Since most stars in the Galaxy are members
of multiple systems, he concludes that collapsing cloud cores must
generally arise from configurations less steep than $\rho \propto
r^{-2}$.  This point of view is supported by Ward-Thompson et al.
(1994), who claim that observed prestellar molecular cloud cores always
have substantial central portions that are flat, $\rho \approx$ const,
rather than continue along the power law, $\rho \propto r^{-2}$, that
characterizes their outer regions.  It should be noted, however, that
such configurations are in fact consistent with the predictions of
theoretical calculations of molecular-cloud core-evolution by ambipolar
diffusion (Nakano 1979, Lizano \& Shu 1989, Basu \& Mouschovias 1994),
which show that nearly pure power-laws, $\rho \propto r^{-2}$, arise
only for a single instant in time, the pivotal state (Li \& Shu 1996),
just before the onset of protostar formation by dynamical infall.
Moreover, more recent analyses of the millimeter- and
submillimeter-wave dust-emission profiles by Evans et al. (2000) and
Zucconi \& Walmsley (2000) that take into account the drop in dust
temperature (but perhaps, not the gas temperature) in the central
regions of externally irradiated dark clouds show that the portion of
the density profile of prestellar cloud cores that is flat ($\rho
\approx$ const), if present at all, is considerably smaller than
originally estimated by Ward-Thompson et al. (1994).

One can also note that while the Taurus molecular-cloud region
represents the classic case of isolated star formation
(Myers \& Benson 1983), it
contains, if anything, more than its cosmic share of binaries (Ghez,
Neugebauer \& Matthews 1993; Leinert et al. 1993; Mathieu 1994; Simon
et al. 1995; Brandner et al. 1996).  Moreover, when observed by
radio-interferometric techniques, Taurus contains many cloud cores that
are well fit by $\rho \propto r^{-2}$ envelopes, yet each star-forming
core typically contains {\it multiple} young stellar objects (Looney,
Mundy \& Welch 1997).

Recent high-resolution simulations of the fragmentation problem carried
out with a\-dap\-ti\-ve-mesh techniques (Truelove et al. 1998) indicate that
many of the previous hydrodynamical simulations claiming successful
fragmentation with density laws less steep than $\rho \propto r^{-2}$
contained serious errors.  Indeed, as long as the starting conditions
are smooth and close to being in mechanical equilibrium (i.e., start
with only one Jeans mass), gravitational collapses seem {\it in
general} not to produce fragmentation.  The emphasis on the sole fault
lying with the law $\rho \propto r^{-2}$ is therefore misplaced.
Something else is needed.  Klein et al. (2000) identify the missing
ingredient as cloud turbulence; our opinion is that magnetic fields may
be equally or even more important.

\subsection{The Effect of Magnetic Fields}
\label{sub_mag}

It is a proposition universally acknowledged that on scales larger than
small dense cores, magnetic fields are more important than thermal
pressure (but perhaps not turbulence) in the support of molecular
clouds against their self-gravitation (see the review of Shu, Adams, \&
Lizano 1987).  Mestel has long emphasized that the presence of
dynamically significant levels of magnetic fields changes the
fragmentation problem completely (Mestel \& Spitzer~1956; Mestel~1965a,b;
Mestel~1985).  Associated with the flux $\Phi$ frozen into a cloud (or any
piece of a cloud) is a magnetic critical mass:  
\be 
M_{\rm cr}(\Phi) = {\Phi \over 2\pi G^{1/2}}.  
\label{MPhi} 
\ee 
Subcritical clouds with masses $M$ less than $M_{\rm cr}$ have magnetic
(tension) forces that are generally larger than and in opposition to
self-gravitation (e.g., Shu \& Li 1997) and cannot be induced to
collapse by any increase of the external pressure.  Supercritical
clouds with $M > M_{\rm cr}$ do have the analog of the Jeans mass -- or,
more properly, the Bonnor-Ebert mass -- definable for them, but unless
they are highly supercritical, $M\gg M_{\rm cr}$, they do not easily
fragment upon gravitational contraction.  The reason is that if $M\sim
M_{\rm cr}$ for the cloud as a whole, then any piece of it is likely to be
subcritical since the attached mass of the piece scales as its volume,
whereas the attached flux scales as its cross-sectional area.  Indeed,
the piece remains subcritical for any amount of contraction of the
system, as long as the assumption of field freezing applies.  An
exception holds if the cloud is highly flattened, in which case the
enclosed mass and enclosed flux of smaller pieces both scale as the
cross-sectional area. This observation led Mestel (1965, 1985) to
speculate that isothermal supercritical clouds, upon contraction into
highly flattened objects, could and would gravitationally fragment.
The present paper casts doubt on this speculation (a) when the original
cloud begins from a state of mechanical equilibrium, and (b) when
magnetic flux is conserved by the contracting cloud (see also Shu \& Li
1997).

Zeeman observations of numerous regions (see the summary by Crutcher
1999) indicate that molecular clouds are, at best, only marginally
supercritical.  The result may be easily justified after the fact as a
selection bias (Shu et al.  1999).  Highly supercritical clouds have
evidently long ago collapsed into stars; they are not found in the Galaxy
today.  Highly subcritical clouds are not self-gravitating regions;
they must be held in by external pressure (or by converging fluid
motions); thus, they do not constitute the star-forming
molecular-clouds that are candidates for the Zeeman measurements
summarized by Crutcher (1999).  The clouds (and cloud cores) of
interest for star formation today are, by this line of reasoning,
marginally supercritical almost by default.

The above comments motivate our interest in re-examining the entire
question of binary-star formation by the fission and fragmentation
mechanisms, but including the all-important dynamical effects of
magnetic fields and the empirically well-founded assumption that
pre-collapse cloud cores have radial density profiles that, in first
approximation, can be taken as $\rho\propto r^{-2}$.  Li \& Shu (1996;
see also Baureis, Ebert \& Schmitz~1989) have shown that the general,
axisymmetric, magnetized equilibria representing such pivotal states
assume the form of singular isothermal toroids (SITs):  $\rho(r,\theta)
\propto r^{-2}R(\theta)$ in spherical polar coordinates
$(r,\theta,\varphi)$, where $R(\theta) = 0$ for $\theta =0$ and $\pi$
(i.e., the density vanishes along the magnetic poles).  We regard these
equilibria as the isothermal (rather than incompressible) analogs of
Maclaurin spheroids, but with the flattening produced by magnetic
fields rather than by rotation.  In the limit of vanishing magnetic
support, SITs become singular isothermal spheres (SISs).  In the limit
where magnetic support is infinitely more important than isothermal gas
pressure, SITs become singular isothermal disks (SIDs), with $\rho
(\varpi,z) = \Sigma (\varpi) \delta (z)$ in cylindrical coordinates
$(\varpi,\varphi,z)$, where $\delta (z)$ is the Dirac delta function,
and the surface density $\Sigma (\varpi) \propto \varpi^{-1}$.

In a fashion analogous to the SIS (Shu 1977), the gravitational
collapses of SITs have elegant self-similar properties (Allen \& Shu
2000).  But it should be clear that the formation of binary and
multiple stars could never result from any calculation that imposes a
priori an assumption of axial symmetry.  In this regard, we would do
well to remember the warning of Jacobi in 1834:

\noindent
``One would make a grave mistake if one supposed that the axisymmetric
spheroids of revolution are the only admissible figures of equilibrium.''

Motivated by the insights of those who have preceded us, we therefore
start the campaign to  understand binary and multiple star-formation by
considering in this paper the nonaxisymmetric equilibria of
self-gravitating, magnetized, differentially-rotating, completely
flattened SIDs, with critical or supercritical ratios of mass-to-flux
in units of $(2\pi G^{1/2})^{-1}$, 
\be
\lambda\equiv 2\pi G^{1/2} {M(\Phi)\over \Phi},
\label{deflambda}
\ee
with $\lambda \ge 1$ (see Li \& Shu 1996, Shu \& Li~1997).  Keeping
$\lambda$ fixed, i.e., under the assumption of field freezing, we shall
find that such sequences of non-axisymmetric SIDs bifurcate from their
axisymmetric counterparts at the analog of the dimensionless squared
rotation rate $\beta$ (which we denote in our problem as $D^2$) given
by the linearized stability analysis of Paper I (Shu et al. 2000; see
also Syer \& Tremaine 1996).  Although some of these (Dedekind-like)
sequences produce buds that look as if they might separate into two or
more bodies, we find that, before the separation can be completed (by
secular evolution?), the sequences terminate in shockwaves that
transport angular momentum outward and mass inward in such a fashion as
to prevent fission.

In a future study, we shall follow the gravitational collapse of some
of these non-axisymmetric pivotal SIDs.  The linearized stability
analysis and nonlinear simulations of Paper I suggests that the
collapse of gravitationally unstable axisymmetric SIDs lead to
configurations that are stable to further collapse but dynamically
unstable to an infinity of nonaxisymmetric spiral modes that again
transport angular momentum outward and mass inward in such a fashion as
to prevent disk fragmentation.  We suspect the same fate awaits the
collapse of pivotal SIDs that are non-axisymmetric to begin with, as
long as we continue with the assumption of field freezing.  Thus, we
shall speculate that {\it rapid (i.e., dynamical rather than
quasi-static) flux loss} during some stage of the star formation
process is an essential ingredient to the process of gravitational
fragmentation to form binary and multiple stars from present-day
molecular clouds.

The rest of this paper is organized as follows.  In \S \ref{sec_SID} we
derive the general equations governing the equilibrium of magnetized,
scale-free, non-axisymmetric, self-gravitating SIDs with uniform
velocity fields. In \S \ref{sec_stat} we show that for SIDs with no
internal motions the equations of the problem can be solved
analytically. For the more general case, in \S \ref{sec_lin} we present
an analytical treatment of the slightly nonlinear regime, when
deviations from axisymmetry are small, valid for arbitrary values of the
internal velocity field.  In \S \ref{sec_Mfold} we describe a numerical
scheme to compute non-axisymmetric SIDs for arbitrary values of the
parameters of the problem.  Finally, in \S \ref{sumdisc} we summarize
the implications of our findings for a viable theory of binary and
multiple star-formation from the gravitational collapse of
supercritical molecular cloud cores that start out in a pivotal state
of unstable mechanical equilibrium.

\section{Magnetized Singular Isothermal Disks}
\label{sec_SID}

The governing equations of our problem are given in Paper I.  They are
the usual gas dynamical equations for a completely flattened disk
confined to the plane $z=0$ except for two modifications introduced by
the presence of magnetic fields that thread vertically through the
disk, and that fan out above and below it without returning back to the
disk.

The theorems of Shu \& Li (1997) for isopedically magnetized disks are
valid for arbitrary distribution of the surface density in the disk and
under the hypothesis of magnetostatic equilibrium in the vertical
direction.  First, magnetic tension reduces the effective gravitational
constant by a multiplicative factor $\epsilon \le 1$, where
\be
\epsilon = 1-{1\over \lambda^2}, 
\label{defeps}
\ee
with the dimensionless mass-to-flux ratio $\lambda\ge 1$ taken to be a
constant both spatially (the isopedic assumption) and temporally (the
field-freezing assumption).  Second, the gas pressure is augmented by
the presence of magnetic pressure; this increases the square of the
effective sound speed by a multiplicative factor $\Theta \ge 1$, where
we follow Paper I in adopting
\be
\Theta = {\lambda^2+3\over \lambda^2 + 1}.
\label{deftheta}
\ee

\subsection{Equations for Steady Flow}
\label{sub_eq}

Consider the time-independent equation of continuity in 2D:
\be
\nabla \cdot (\Sigma {\bf u}) = 0.
\ee
This equation can be trivially satisified by adopting a streamfunction 
$\Psi$ defined by
\be
\Sigma {\bf u} = \nabla \times (\Psi \hat {\bf e}_z),
\ee
which written in cylindrical polar coordinates reads
\be
u_\varpi = {1\over \varpi \Sigma}{\partial \Psi
\over \partial \varphi}, \qquad u_\varphi =
-{1\over \Sigma}{\partial \Psi\over \partial \varpi}. 
\label{vel}
\ee
Notice that ${\bf u}\cdot \nabla \Psi = 0$, so curves of constant $\Psi$
describe streamlines.

The momentum equation along streamlines can be replaced by Bernoulli's 
theorem:
\be
{1\over 2}|{\bf u}|^2 + \Theta {\cal H}(\Sigma) +\epsilon {\cal V} = {\cal B}(\Psi), 
\label{along}
\ee
where ${\cal B}(\Psi)$ is the Bernoulli function and
${\cal H}(\Sigma)$ is the specific enthalpy associated
with a barotropic equation of state (EOS) for the gas alone:
\be
{\cal H}(\Sigma) \equiv \int_0^{\Sigma} {d\Pi\over d\Sigma}\, 
{d\Sigma\over\Sigma}.
\label{enth}
\ee
In equation (\ref{enth}) the vertically intgrated pressure $\Pi$ is
assumed to be a function of surface density $\Sigma$ alone.  For an
isothermal EOS, we have $\Pi = a^2\Sigma$ with $a^2=$ const, so that
${\cal H} = a^2 \ln \Sigma$ plus an arbitrary additive constant that we are
free to specify for calculational convenience.

In terms of the variables introduced above,
the vector momentum equation can now be written
\be
(\nabla \times {\bf u})\times {\bf u} + {\cal B}^\prime
(\Psi)\nabla \Psi = 0.
\ee
Expressed in component form, this equation gives the additional
independent relation for momentum balance across streamlines:
\be
{1\over \varpi}\left[ {\partial \over \partial \varpi}
\left( {\varpi\over \Sigma}{\partial \Psi \over \partial
\varpi}\right) +{1\over \varpi}{\partial \over \partial
\varphi}\left( {1\over \Sigma}{\partial \Psi\over \partial
\varphi}\right)\right] = \Sigma {\cal B}^\prime (\Psi) .
\label{across}
\ee
Notice that the LHS is the $z$-component of $-\nabla \times {\bf u}$;
thus, $\Sigma {\cal B}^\prime$ is the local vorticity contained in the
flow (proportional to Oort's $B$ constant).  The above set of equations
is closed by the addition of Poisson's equation:
\be
{\cal V}(\varpi,\varphi) = -G\oint \,d\psi
\int_0^\infty {\Sigma (r,\psi) \,rdr\over
[r^2+\varpi^2-2r\varpi \cos (\psi-\varphi)]^{1/2}} .
\label{poisson}
\ee

\subsection{Scale-Free Isothermal Solutions}
\label{sub_scale}

For aligned SIDs, we look for solutions of the form,
\be
{\cal H}(\Sigma) = a^2\lim_{R\rightarrow \infty}
\ln \left({2R\Sigma\over K}\right), 
\label{defH}
\ee
\be
\Sigma (\varpi,\varphi) = {K\over \varpi}S(\varphi), 
\label{dens}
\ee
where the constant $K$ with dimension of g cm$^{-1}$ and the
dimensionless function $S(\varphi)$ are to be determined.  In equation
(\ref{defH}) and in everything that follows, the limit operation
$R\rightarrow \infty$ is to be taken after differentiation of variables
like $\cal H$ and $\cal V$ in the equations of motion have occurred.
We have taken
advantage of the fact that additive constants in variables like $\cal
H$, $\cal V$, and $\Psi$ do not enter the physical equations of motion
to introduce a temporary artificial radial scale $R$ so that we need
not take logarithms of dimensional quantities.  Putting the freedom to
scale $\Sigma$ entirely into $K$, we are free to normalize the function
$S(\varphi)$ such that
\be
{1\over 2\pi}\oint S(\varphi)\, d\varphi \equiv 1. 
\label{norm}
\ee

Substitution of equation (\ref{dens}) into equation (\ref{poisson}) yields
\be
{\cal V}(\varpi,\varphi)=-GK \lim_{R\rightarrow \infty} \oint S(\psi)\,d\psi
\int_0^{R/\varpi} {dx\over [1-x^2-2x\cos (\varphi-\psi)]^{1/2}},
\label{poisson1}
\ee
where $x\equiv r/\varpi$. The inner integral can be evaluated by
elementary techniques and gives
\be
\ln \{(R/\varpi) -\cos (\varphi-\psi) +[1+ (R/\varpi)^2 -2(R/\varpi)
\cos(\varphi-\psi)]^{1/2}\} -\ln [1-\cos(\varphi-\psi)].
\ee
The argument of the first logarithm equals
\be
(R/\varpi)\left\{ 1-(\varpi/R)\cos(\varphi-\psi)+\left[1-2(\varpi/R)
\cos(\varphi-\psi)+(\varpi/R)^2\right]^{1/2}\right\},
\ee
which can be expanded for large $R$ as
\be
(2R/\varpi)[ 1 - (\varpi/R)\cos(\varphi-\psi) +\dots].
\ee
Thus, the inner integral in equation (\ref{poisson1}) equals
\be
\ln(2R/\varpi) + \ln[1-(\varpi/R)\cos(\varphi-\psi)+\dots]-
\ln[1-\cos(\varphi-\psi)].
\ee
For large $R$, the middle logarithm goes to zero, and
the substitution of the above result then yields
\be
{\cal V}(\varpi,\varphi) = 2\pi GK \lim_{R\rightarrow \infty} \left[
\ln (\varpi/2R) + V(\varphi)\right], 
\label{gravpot}
\ee
where
\be
V(\varphi) = {1\over 2\pi}\oint S(\psi)\ln[1-\cos(\psi-\varphi)]\, d\psi . 
\label{defV}
\ee

We further look for solutions of the form,
\be
\Psi(\varpi,\varphi) =\Theta^{1/2}a K \lim_{R\rightarrow
\infty} [ -D\ln (\varpi/2R) + W(\varphi)], 
\label{defPsi}
\ee
\be
{\cal B}(\Psi) = - \Theta^{1/2}a {B\over K} \Psi,
\label{calB}
\ee
where $D$ and $B$ are dimensionless constants whose values are yet to be
specified.  In what follows, it is convenient to define the dimensionless
radial mass flux as
\be
U(\varphi) \equiv W^\prime (\varphi), 
\label{defU}
\ee
which we will regard as an ODE for the angular part of the streamfunction
$W(\varphi)$ if we know $U(\varphi)$.  An integration of equation (\ref{defU})
over a complete cycle shows that the mass flow across a full circle
must vanish,
\be
\oint U(\varphi) \, d\varphi = 0, 
\label{Uint}
\ee
since $W(\varphi)$ is a periodic function of $\varphi$.  In order for
equation (\ref{Uint}) to hold nontrivially, $U(\varphi)$ must possess both
positive and negative values; thus, it must pass through zero at least
once in the range $(-\pi,+\pi)$.  We define our angular coordinate so
that $U(\varphi)$ is zero at $\varphi = 0$:
\be
U(0) = 0. 
\label{Ubc}
\ee
This convention results in $U(\varphi)$ being an odd function of
$\varphi$.

Substitution of the expression for $\Sigma$ and $\Psi$ into
equation (\ref{vel}) now yields the identifications:
\be
u_\varpi = \Theta^{1/2}a {U(\varphi)\over S(\varphi)}, \qquad
u_\varphi = \Theta^{1/2}a {D\over S(\varphi)}.
\label{ucomp}
\ee
In other words, apart from the compression and decompression factor
$S(\varphi)$ as fluid elements flow in azimuth in a nonaxisymmetric
disk, the dimensionless function $U(\varphi)$ is the generator for
radial motions and the dimensionless constant $D$ is the generator for
angular motions.

The substitution of equations (\ref{enth}), (\ref{dens}), (\ref{defV}),
(\ref{defPsi}), (\ref{calB}), and (\ref{ucomp}) into equation
(\ref{along}) now yields $K$ from the the radial part of the equality,
\be
K = {\Theta a^2\over 2\pi \epsilon G}(1+DB),
\label{defK}
\ee
whereas the angular part of the equality gives
\be
{1\over 2S^2}\left(U^2 + D^2\right) + (1+DB)V + \ln S = -BW.
\label{bernoulli}
\ee
Similarly, equation (\ref{across}) leads to the requirement,
\be
-{D\over S} + {d\over d\varphi}\left( {U\over S}\right) = - BS.
\label{gs}
\ee
Since the combination $U/S$ must be a periodic function of
$\varphi$, we may integrate equation (\ref{gs}) over a complete
cycle and obtain the further constraint:
\be
B = {D\over 2\pi}\oint {d\varphi \over S(\varphi)}.
\label{defB}
\ee
Finally, differentiating equation (\ref{bernoulli}) with respect to $\varphi$ 
and using equation (\ref{gs}) we obtain 
\be
(S^2-D^2)S^\prime+DUS+(1+DB)S^3 V^\prime=0. 
\label{bernoulli1}
\ee
Equation (\ref{bernoulli1}) possesses critical points at $S(\varphi)=D$,
where $u_\varphi$ becomes equal to the magnetosonic 
speed (see eq.~\ref{ucomp}).

Equations (\ref{defV}), (\ref{gs}), (\ref{defB}) and (\ref{bernoulli1})
are the fundamental set of integro-differential equations governing the
problem. They have to be solved in the interval $\varphi=[0,2\pi]$ for
the three unknown functions $S(\varphi)$, $V(\varphi)$, $U(\varphi)$
and the unknown constant $B$. The constant $D$ itself is freely
specifiable.  Notice that the arbitrarily introduced radial scale $R$
enters nowhere in the final equations.

Notice also that equation (\ref{gs}) implies that radial motions arise
only in response to a local imbalance of forces -- gravitational,
pressure, and inertial -- across streamlines, even though equation
(\ref{defB}) requires such forces to be balanced on average over a
circle.  Moreover, the governing equations (\ref{defV}), (\ref{gs}),
and (\ref{bernoulli1}) require $S(\varphi)$ and $V(\varphi)$ to be
symmetric with respect to $\varphi=\pi$ when $U(\varphi)$ is chosen to
be antisymmetric.  In other words, $\Sigma(\varphi)$ and $V(\varphi)$ are
cosine series in $\varphi$ when $U(\varphi)$ is developed as a sine
series. Consequently, the choice of the zero of the angular
coordinate is not unique: for a configuration with a basic $M$-fold
symmetry, where $M$ is a positive integer, the condition
$U(\varphi)=0$ is satisfied in the interval $[0,2\pi]$ at $\varphi=
k\pi/M$ with $k=0,1,2,\ldots,2M$, and different choices of the 
$x$-axis correspond to rotations of the equilibrium configuration 
by multiples of $\pi/M$.

\subsection{Fourier Decomposition for Poisson Integral}
\label{sub_dec}

When we have departures from axial symmetry, the difference form of the
kernel and the periodic nature of the wanted solutions makes equation
(\ref{defV}) suitable for solution by Fourier series.  Since $S(\varphi)$
and $V(\varphi)$ are periodic functions of $\varphi$, they are expandable
as the series:
\be
S(\varphi) = S_0+\sum_{m=1}^\infty S_m \cos (m\varphi),
\label{Sfou}
\ee
\be
V(\varphi) = V_0+\sum_{m=1}^\infty V_m \cos (m\varphi), 
\label{Vfou}
\ee
where we have isolated the axisymmetric terms $S_0$ and $V_0$.
The coefficients $S_m$ and $V_m$ are real for all $m\ge 1$.
In writing $S(\varphi)$ and $V(\varphi)$ as pure cosine series, we have
made use of our freedom to orient one of the principal figure axes of
aligned equilibria along the $x$-axis.

Since equation (\ref{defV}) gives $V(\varphi)$ as a convolution of
$S(\varphi)$ and $\ln (1-\cos\varphi)$, substitution of equations
(\ref{Sfou}) and (\ref{Vfou}) into equation (\ref{defV}) and application
of the convolution theorem for Fourier cosine transforms result in the
identification,
\be
V_m = L_m S_m,
\ee
where
\be
L_m \equiv {1\over 2\pi}\oint \ln (1-\cos \varphi)
\cos(m\varphi) \, d\varphi=
\left\{ 
\begin{array}{ll}
-\ln 2 & \mbox{if $m=0$}       \\
-1/|m| & \mbox{if $|m| \ge 1$},
\end{array}
\right.
\ee
as shown in the Appendix. 
Therefore, the normalization condition (\ref{norm}) implies 
\be
S_0=1, \qquad V_0=-\ln 2,
\ee
and equations (\ref{Sfou}) and (\ref{Vfou}) can be written as
\be
S(\varphi) = 1-\sum_{m=1}^\infty mV_m \cos (m\varphi),
\label{Sfou1}
\ee
\be
V(\varphi) = -\ln 2+\sum_{m=1}^\infty V_m \cos (m\varphi).
\label{Vfou1}
\ee

If we are given $\{V_m\}_{m=1}^\infty$, then we know $S(\varphi)$
and $V(\varphi)$.  Unfortunately, local knowledge of either $S$ or $V$
at $\varphi$ does not determine the value of the other at the same
$\varphi$.  The relationship is local in Fourier space, so only global
knowledge of $S(\varphi)$ gives global knowledge of $V(\varphi)$; i.e.,
$S(\varphi)$ is a functional, and not a function, of $V(\varphi)$.

Notice that in general, if a set $V_m$ of Fourier coefficients
corresponds to a solution, then the set $(-1)^m V_m$ corresponds to the
same configuration rotated by an angle $\pi$, as discussed at the end
of \S \ref{sub_scale}.

\section{Static Equilibria}
\label{sec_stat}

For static equilibria, $U=D=0$ and equation (\ref{bernoulli1}) reduces to 
\be
S^\prime+SV^\prime=0,
\label{bernoulli2}
\ee
which has the barometric solution: $S(\varphi) = Ae^{-V(\varphi)}$,
where $A$ is a constant that can be adjusted to satisfy the normalization
condition (\ref{norm}).   Substitution of $S(\varphi) =
Ae^{-V(\varphi)}$ into Poisson's integral (eq.~\ref{defV}), or alternatively
its Fourier decomposition (eq.~\ref{Sfou1} and \ref{Vfou1}), then
constrains the solution for $V(\varphi)$.
Remarkably, this system of nonlinear functional relations has an 
analytical solution, where iso-surface-density contours are {\it ellipses} of
eccentricity $e$, 
\be
S(\varphi)={\sqrt{1-e^2}\over 1\pm e\cos\varphi},
\label{S1}
\ee
with $0<e<1$, the $\pm$ sign representing our freedom to rotate 
the equilibrium configuration by an angle $\pi$, as anticipated 
at the end of \S\ref{sub_scale}. Let us consider the case 
with the minus sign (the proof for the case with the plus sign 
is completely analogous).

The set of Fourier coefficients corresponding to equation (\ref{S1}) is
\be 
V_m=-{1\over \pi m}\oint S(\varphi)\cos(m\varphi)\, d\varphi=
-{2\sqrt{1-e^2}\over \pi m}
\int_0^\pi{\cos(m\varphi)\over 1-e\cos\varphi}\, d\varphi=
-{2\over m}\left({1-\sqrt{1-e^2}\over e}\right)^m,
\ee
where we have used formula (3.613.1) of Gradshteyn \& Ryzhik (1965) to
evaluate the integral.
From equation (\ref{Vfou1}) we obtain the potential
\be
V(\varphi)=-\ln 2-2\sum_{m=1}^\infty \left({1-\sqrt{1-e^2}\over e}\right)^m 
{\cos(m\varphi)\over m} =\ln{1-e\cos\varphi\over 1+\sqrt{1-e^2}},
\label{V1}
\ee
where we have evaluated the sum of the series by using formula (1.448.2)
of Gradshteyn \& Ryzhik (1965).
By direct substitution, therefore, we see that $S$ and $V$
are related by the barometric relationship implied
by equation (\ref{bernoulli2}): $S=Ae^{-V}$,
where $A = \sqrt{1-e^2}/(1+\sqrt{1-e^2})$.  (QED)

The static axisymmetric solution (a magnetized but nonrotating disk
with surface density $\Sigma=K/\varpi$) is trivially recovered setting
$e=0$; Li \& Shu (1997) give the time-dependent self-similar
gravitational collapse of this special case.   In the other extreme,
for $e=1$ the potential becomes
\be
V(\varphi)=\ln(1-\cos\varphi),
\label{sing2}
\ee
and the corresponding set of Fourier coefficients, $V_m=-2/m$, 
substituted into equation (\ref{Sfou1}),
gives the familiar Fourier expansion of 
the Dirac $\delta$-function, 
\be
S(\varphi) = 1+2\sum_{m=1}^\infty\cos(m\varphi)\equiv 2\pi\delta(\varphi).
\label{sing1}
\ee
That the limit $e\rightarrow 1$ produces a semi-infinite
filament with mass per unit length $2\pi K$
also follows from equation (\ref{S1}).
For values of $e$ between these two extremes, both iso-surface-density
contours and equipotentials are confocal ellipses of eccentricity $e$.
Figure 1 shows some examples of static SIDs for different values of 
the eccentricity $e$.

With our $x$-axis relabelled as the $z$ axis, our filament solution is
equivalent to the (nonrotating) combination $A=2$ and $B\rightarrow
\infty$ in the eccentric generalization given by Medvedev \& Narayan
(2000) of the axisymmetric singular isothermal toroids found by Toomre
(1982) and Hayashi, Narita, \& Miyama (1982).  We differ from Medvedev
\& Narayan (2000), however, in the opinion whether such lopsided
configurations represent legitimate states of equilibrium.  Unlike the
situation for systems with greater angular symmetry, the gravitational
field of eccentric distributions of matter does not vanish at the
origin.  Nevertheless, in both the SID and SIT solutions (static or
rotating), the (infinite) gravitational force at the origin is exactly
balanced by an (infinite) pressure gradient acting in concert perhaps
with an (infinite) centrifugal force.  This balance is qualitatively no
different than at any other point in the system, and it would be an
artificial restriction to rule out eccentric equilibria simply because
they have a nontrivial balance of forces at the origin rather than a
trivial one.  The worry by Toomre cited by Medvedev and Narayan that
eccentric SITs (and SIDs presumably) are {\it unstable} equilibria is a
different matter, and may be overshadowed, at least for gaseous
configurations, by the knowledge that the more slowly rotating members
of singular isothermal equilibria, whether lopsided or not, are all
unstable to inside-out gravitational collapse similar to the familiar
case of the SIS (Shu 1977), whereas the more rapidly rotating members
are prone to spiral instabilities via swing amplification (Toomre 1977;
see also Paper I).

\subsection{A Specific Example: the Molecular Cloud Core L1544}
\label{sub_L1544}

As an amusing sideshow, Figure 2 shows an overlay of one of our
eccentrically displaced static models projected onto a map of thermal
dust emission at 1.3~mm obtained by Ward-Thompson, Motte, \& Andr\'e
(1999) for the prestellar molecular cloud core L1544.  Apart from
relatively minor fluctuations due to the cloud turbulence, the solid
curves depicting the iso-surface-density contours of the theoretical
model match well both the observed shapes and
grey-scale of the dust isophotes.

Zeeman measurements of the magnetic-field component parallel to our
line of sight toward L1544 have been made by Crutcher \& Troland
(2000), who obtain $B_\parallel = 11\pm 2$ $\mu$G.  For a highly
flattened disk, which is reflection symmetric about the plane $z=0$,
integration along the line of sight yields cancelling contributions of
$B_\varpi$ and $B_\varphi$ to $B_\parallel$. The $z$-component of the
magnetic field of our model core is given by
\be
B_z={2\pi G^{1/2}\over \lambda} \Sigma.
\ee
We may now calculate the average value of $\Sigma$ within a radius $R$ as
\be
\langle \Sigma \rangle= {1\over \pi R^2}
\oint d\varphi \int_0^R \Sigma \varpi\, d\varpi=
{\Theta a^2 \over \pi \epsilon G R} 
= {\lambda^2(\lambda^2+3)\over (\lambda^4-1)}
{a^2\over \pi G R},
\ee
where we have made use of equations (\ref{defeps}), (\ref{deftheta}),
(\ref{dens}), (\ref{norm}) and ({\ref{defK}). Therefore, the average 
value of $B_z$ within a radius $R$ is
\be
\langle B_z \rangle = {2\pi G^{1/2}\over \lambda}\langle \Sigma\rangle
= {\lambda (\lambda^2+3)\over (\lambda^4-1)} {2 a^2\over G^{1/2} R}.
\ee
Notice the pleasant result that the above formulae do not involve $e$.

Since we model L1544 as a thin disk with elliptical iso-surface-density
contours, its orientation in space is defined by three angles, two
specifying the orientation of the disk plane, the third giving the
position of the elliptical contours in this plane. We fix the first
angle by assuming for simplicity that the major axis of the elliptical
contours lies in the plane of the sky.  The second angle $i$ is the
inclination of the minor axis with respect to the plane of the sky
($i=0$ for a face-on disk) and can be adjusted to fit the
observations.  The third angle, specifying the ellipse's
orientation in the disk plane, is given as 38$^\circ$
north through east by Ward-Thompson et al.~(2000).

We choose the eccentricity $e$ and inclination $i$ by the
following procedure.  From Figure 2, we can
estimate that a typical dust contour has a ratio
of distances closest and farthest from the core center
given in a model of nested confocal ellipses by
\be
{1-e\over 1+e} \approx 0.30 \qquad \Rightarrow \qquad e \approx 0.54.
\ee
Similarly, we may estimate that these ellipses have
an apparent minor-to-major axis-ratio of
\be
(1-e^2)^{1/2}\cos i \approx 0.54 \qquad \Rightarrow
\qquad \cos i \approx 0.64.
\ee
The resulting ellipses for three
iso-surface-density contours, spaced in a geometric progression
1:2:4, are shown as solid curves in Figure 2.

Determination of $\cos i$ allows us to compute an expected
$\langle B_\parallel\rangle$ = $\langle B_z \rangle\cos i$.  Similarly,
we obtain the expected hydrogen column density
by multiplying $\langle \Sigma \rangle$ by $(\cos i)^{-1}$
for a slant path through an inclined sheet
and by 0.7 for the mass fraction of H nuclei of mass $m_H$:
$N_H=0.7 \langle \Sigma \rangle / (m_H\cos i)$.

The sound speed for the 10 K gas in L1544 is $a=0.19$~km~s$^{-1}$
(Tafalla et al.~1998).  These authors give $\Delta V = 0.22$ km
s$^{-1}$ as the typical linewidth for their observations of C$^{34}$S
in this region.  For such a heavy molecule, turbulence is the main
contributor to the linewidth, which allows us to estimate the mean
square turbulent velocity along a typical direction (e.g., the line of
sight) as $v_t^2 = \Delta V^2/8\ln 2$.  We easily compute that $v_t^2$
has only 24\% the value of $a^2$.  Assuming that it is possible to
account for the ``pressure'' effects of such weak turbulence by adding
the associated velocities in quadrature, $a^2+v_t^2$, we adopt an
effective isothermal sound speed of $a=0.21$ km s$^{-1}$ for L1544.

The radius of the Arecibo telescope beam at the distance of L1544 is
$R=0.06$~pc (Crutcher \& Troland~2000).  Ambipolar diffusion
calculations by Nakano (1979), Lizano \& Shu (1989), Basu \&
Mouschovias (1994) suggest that $\lambda \approx 2$ when the pivotal
state is approached (see the summary of Li \& Shu 1996).  Putting
together the numbers, $\cos i = 0.64$, $R = 0.06$ pc, $a = $ 0.21 km
s$^{-1}$, and $\lambda = 2$, we get $\langle
B_\parallel\rangle=11$~$\mu G$, in excellent agreement with the Zeeman
measurement of Crutcher \& Troland (2000).  These authors also deduce
$N_H=1.8\times 10^{22}$ cm$^{-2}$ from their OH measurements, whereas
we compute a hydrogen column density within the Arecibo beam of
$N_H=1.4 \times 10^{22}$ cm$^{-2}$.  The slight level of disagreement
is probably within the uncertainties in the calibration or calculation
of the fractional abundance of OH in dark clouds (cf. Crutcher 1979,
van Dishoeck \& Black 1986, Flower 1990, Heiles et al. 1993).

Our ability to obtain good fits of much of the observational data
concerning the prestellar core L1544 with a simple analytical model
should be contrasted with other, more elaborate, efforts.  Consider,
for example, the {\it axisymmetric} numerical simulation of Ciolek \&
Basu~(2000), who assumed a disk close to being edge-on ($\cos i \approx
0.3$ when $e$ is assumed to be 0) to reproduce the observed elongation,
but who left unexplained the eccentric displacement of the cloud core's
center (very substantial for ellipses of eccentricity $e \approx
0.54$).   The adoption of axisymmetric cores leads to another problem:
Ciolek \& Basu's deprojected magnetic field is on average 3-4 times
stronger than ours, values probably in conflict with Zeeman
measurements of low-mass cloud cores.  [See the comments of Crutcher \&
Troland (2000) concerning the need for magnetic fields in Taurus to be
all nearly in the plane of the sky if conventional models are
correct.]  Natural elongation plus projection effects, as anticipated
in the comments of Shu et al.~(1999), allow us to model L1544 as a
moderately supercritical cloud, with $\lambda \approx 2$, fully
consistent with the theoretical expectations from ambipolar diffusion
calculations, and in contrast with the value $\lambda \approx 8$
estimated by Crutcher \& Troland~(2000) from the measured values of
$B_\parallel$ and $N_H$. In addition, if L1544 is a thin, {\it
intrinsically eccentric}, disk seen moderately face-on, as implied by
our model, then the extended inward motions observed by Tafalla et
al.~(1998; see also Williams et al.~1999) may be attributable to a
(relatively fast) core-amplification mechanism that gathers gas
(neutral and ionized) dynamically but subsonically along magnetic field
lines on both sides of the cloud toward the disk's midplane.

Finally, we show in Figure 2 the direction of the average magnetic
field projected in the plane of the sky predicted by our model (thin solid
line) and derived from submillimeter polarization observations of
Ward-Thompson et al.~(2000) (thin dashed line). Since we have assumed
in our model that the major axis of iso-surface-density contours is in
the plane of the sky, the predicted projection of the magnetic field is
parallel to the cloud's minor axis. The offset between the measured
position angle of the magnetic field and the cloud's minor axis might
indicate some inclination of the cloud's major axis with respect to the
plane of the sky.  The turbulent component of the magnetic field, not
included in our model, may also contribute to the observed deviation.

\section{Linear Perturbations of Axisymmetric Rotating SIDs}
\label{sec_lin}

We now consider equilibrium configurations with internal motions:
$D\neq 0$, $U(\varphi)\neq 0$.  For comparison with the analysis of
Paper I, we begin with a perturbative analysis of the equations of the
problem valid for small deviation from axisymmetry.

For axisymmetric disks, $V_m=0$ for $m\ge 1$, and therefore equations
(\ref{Sfou1}) and (\ref{Vfou1}) reduce to 
\be
S = S_0= 1, \qquad V =V_0= -\ln 2.
\ee
Iso-surface-density contours are now circles.
Substitution of these values into equation (\ref{defB}) 
and (\ref{bernoulli1}) yields $B=D$ and $U=0$.
The dynamics of centrifugal balance is contained in the relationship
(\ref{defK}) among the various constants of the problem:
\be
K = {\Theta a^2 \over 2\pi \epsilon G}(1+D^2), 
\ee
the same as equation (9) of Paper I. These axisymmetric SIDs are uniquely 
determined, in an irreducible sense, by a freely specifiable value of $D$.
(Physically, we are also free, of course, to choose different scalings
via $a$ and $\lambda$, with the latter determining $\epsilon$ and $\Theta$.)

Consider now small departures from these axisymmetric states 
characterized by a basic $M$-fold symmetry, with $M=1,2,3,\ldots$. 
Equations (\ref{Sfou1}) and (\ref{Vfou1}) give
\be
S(\varphi) = 1-MV_M \cos (M \varphi),
\label{Spert}
\ee
\be
V(\varphi) = -\ln 2 + V_M \cos (M \varphi).
\label{Vpert}
\ee
Equation (\ref{Spert}) shows that for small deviations from
axisymmetric iso-surface-density contours are
{\it lima\c cons of Pascal} (D\"urer 1525).

As required by equation (\ref{gs}),
$U(\varphi)$ must be expanded as a sine series,
\be
U(\varphi) = U_M \sin (M\varphi).
\label{Upert}
\ee
To linear order, $B=D$ as in the axisymmetric case.

Substitution of the relations (\ref{Spert})-(\ref{Upert}) into 
equations (\ref{bernoulli1}) and (\ref{gs}) of the governing set
yields, after subtraction of the axisymmetric relations and linearizing,
\be
M^2(1-D^2)V_M - M(1+D^2)V_M + DU_M=0,
\ee
\be
-DV_M + U_M = DV_M.
\ee
Solutions are possible for arbitrary (infinitesimal) values
of $V_M$ provided
\be
U_M = 2DV_M,
\ee
and
\be
M(1+D^2)-M^2(1-D^2) = 2D^2 . 
\label{linear}
\ee
Equation (\ref{linear}) is equivalent to equation (25)
of Paper I and can be satisfied by $M=1$ for {\it any} rotation rate 
$D$ (including $D$ = 0).  For $M > 1$, we require special values of $D$:
\be
D^2 = {M\over M+2} \qquad {\rm for} \qquad M = 2, 3, 4, \dots
\ee
Notice the result that the required $D^2\rightarrow 1$ as 
$M \rightarrow \infty$.

For any given $D$, different values of $V_M\ll 1$ generate a continuum
of linearized solutions.  Without loss of generality, we can assume
$V_M>0$, as the transformation $V_M\rightarrow -V_M$ is equivalent
to a rotation of the equilibrium configuration by an angle $\pi/M$.
(see discussion at the end of \S \ref{sub_scale}).
To lowest order, the two components of the fluid velocity as given 
by equation (\ref{ucomp}) satisfy
\be
{u_\varpi\over \Theta^{1/2}a}=2DV_M\sin(M\varphi),
\label{urlin}
\ee
and 
\be
{u_\varphi\over\Theta^{1/2}a}=D[1+V_M\cos(M\varphi)].
\label{uplin}
\ee
Therefore, for infinitesimal values of $V_M$ the flow describes 
a locus in the velocity plane $(u_\varpi,u_\varphi)$ which 
is an ellipse of axial ratio 2 centered on $(0,\Theta^{1/2}aD)$:
\be
\left({u_\varpi\over 2\Theta^{1/2}a}\right)^2+
\left({u_\varphi\over \Theta^{1/2}a}-D\right)^2=D^2V_M^2.
\ee
Notice that the axial ratios are a factor of $\sqrt{2}$
larger than the kinematic epicycles a collisionless
body would generate upon being disturbed from a circular
orbit in a disk that has a flat rotation curve (e.g.,
Binney \& Tremaine 1987); the extra factor of $\sqrt{2}$
(and a non-precessing pattern with $M$ lobes) arises
for a {\it fluid} disk because of the coherence enforced
by the collective self-gravity of the perturbations.

As $V_M$ is increased, the flow must eventually try to cross the
magnetosonic point, $u_\varphi=\Theta^{1/2}aD$, which is a singular
point of equation (\ref{bernoulli1}).  This transition cannot be followed
without the introduction of shocks (see the analogous
phenomena of spiral galactic shocks treated by Shu, Milione, \& Roberts
1973).  In the present context, smooth-flow solutions
are possible only if $u_\varphi \le
\Theta^{1/2}a$ (entirely submagnetosonic flow, for $D<1$) or $u_\varphi
\ge \Theta^{1/2}a$ (entirely supermagnetosonic flow, for $D>1$).  When
$D$ is close to 1, either slightly smaller or larger, the azimuthal
velocity in the SID is very close to magnetosonic already in the
axisymmetric case. Thus, the magnetosonic point is reached when
deviations from axisymmetry are small, and the results of the linear
analysis developed above can be applied. Equation (\ref{uplin}) then gives
the critical value of the coefficient $V_M$, in the linear regime, at
which the flow tries to cross the magnetosonic point,
\be
V_M^{\rm crit} \approx \pm\left(1 - {1\over D}\right),
\label{mcrit}
\ee
with the plus (minus) sign valid for $D>1$ ($D<1$).

\section{Fully Nonlinear Models with Internal Motions}
\label{sec_Mfold}

\subsection{Numerical Method}
\label{sec_NumMeth}

In the general case, we solve the set of governing equations by
iteration.  For a given iterate when $S(\varphi)$ is known, we may
regard equation (\ref{defV}) as an integral for $V(\varphi)$.
Similarly, equations (\ref{gs}) and (\ref{Ubc}) constitute an ODE plus
its starting condition for $U(\varphi)$.  For general $\varphi$,
equation (\ref{bernoulli1}) may now be solved as a first order ODE with
the boundary condition equation (\ref{norm}) to obtain a new iterate
for $S(\varphi)$.  The procedure actually adopted substitutes a Fourier
transform for a direct integration of equation (\ref{defV}), as
described in \S \ref{sub_dec}.

\noindent
(A)  Fix the value of $D$ that one wants to study.
Suppose we want to study a configuration with a basic
$M$-fold symmetry, with $M = 1,2,3, \dots$. Then we would begin
with an initial guess for the Fourier coefficients
$\{V_m\}_{m=1}^\infty$. We then compute
\be
V(\varphi) = -\ln 2 + \sum_{m=1}^\infty V_m \cos (mM\varphi),
\ee
and
\be
S(\varphi) = 1 -M\sum_{m=1}^\infty m V_m \cos (mM\varphi).
\ee

\noindent 
(B) Compute the resulting value of $B$ from equation (\ref{defB}).  
Since the cycle
need be taken only over $2\pi/M$ in $\varphi$, we have
\be
B = {MD\over 2\pi}\int_0^{2\pi/M} {d\varphi \over S(\varphi)}.
\ee
Integrate equation (\ref{gs}) for $U$ subject to the starting condition
(\ref{Ubc}).  Since $S$ has been forced to be a cosine series, $U$ is
then automatically a sine series, i.e., we should automatically find
$U(\varphi)$ to be $M$-periodic, with $U(2\pi/M) = 0$.

\noindent
(C)  With $D$ fixed, and with $B$, $V(\varphi)$, $S(\varphi)$, and
$U(\varphi)$, known in the form of the current iterates, solve equation
(\ref{bernoulli1}) as a first order ODE for $S(\varphi)$, subject to the normalization
condition (\ref{norm}).  With this new iterate for $S(\varphi)$ compute the
Fourier coefficients
\be
V_m = -{1\over \pi m}\int_0^{2\pi/M}S(\varphi)\cos (mM\varphi)
\,d\varphi \qquad {\rm for} \qquad m=1,2, \dots
\ee
Compare these coefficients with those from the previous iterate.  If they
are insufficiently precise, go back to step (A), after introducing, if
necessary, a relaxation parameter to smooth between successive iterates
for $V_m$.

\subsection{Numerical Results}
\label{results}

Results from our numerical integrations are illustrated in Figures 3--10. 
It is convenient to define a plane $(D^2, S_M)$, where $S_M=-MV_M$ is the
first coefficient in the Fourier expansion of the function $S(\varphi)$,
and can be considered an indicative measure of deviations from
axial symmetry.  Figure 3 shows the regions in the $(D^2,S_1)$ plane
occupied by $M=1$ models with entirely submagnetosonic or entirely
supermagnetosonic flow.  At the upper limit of these two regions
the flow attempts a magnetosonic transition at perisys (closest
to the system center) in the former case and at aposys
(farthest from the system center) in the latter case, as
computed numerically with the method described in
\S \ref{sec_NumMeth}.  The long-dashed line shows the same
magnetosonic limit as given by equation (\ref{mcrit}) in the linear
approximation $S_1\ll 1$.
Notice that for $D=0$ the results of
\S \ref{sec_stat} show that $S_1^{\rm crit}=V_1^{\rm crit}=2$.   Tick
marks denote the values of $D^2$, as predicted by the linear analysis
of Paper I and \S \ref{sec_lin}, where bifurcations occur with
$M$-fold symmetry $(M \ge 2)$ from the axisymmetric sequence of SIDs
that lie along the short dashed line.

Figure 4 shows submagnetosonic $M=1$ states for the case $D^2=0.1$ as
$S_1$ progresses from the axisymmetric limit ($S_1=0$) to just before
the magnetosonic transition ($S_1=1.39)$.  Notice that flow velocities
are largest at perisys because of the tendency to conserve specific
angular momentum (not exact because the self-consistent gravitational
field is nonaxisymmetric).  As a consequence, the magnetosonic
transition, when it arrives, is made at the minimum of the
gravitational potential, as seen by a fluid element, when the base flow
is submagnetosonic.  Notice also that the iso-surface-density contours
are quasi-elliptical with foci at the center of the system and with the
major axes lying in the same direction as the elongation of the
streamlines formed by connecting the flow arrows.

Figure 5 shows supermagnetosonic $M=1$ states for the case $D^2=4$ as
$S_1$ progresses from the axisymmetric limit ($S_1=0$) to just before
the magnetosonic transition ($S_1=1.08)$.  Notice that flow velocities
are smallest at aposys, again because of the (inexact) tendency to
conserve specific angular momentum.  As a consequence, the magnetosonic
transition, when it arrives, is made at the maximum of the
gravitational potential, as seen by a fluid element, when the base flow
is supermagnetosonic.  Notice also that the iso-surface-density
contours are now elongated in the opposite sense to streamlines made by
connecting the flow arrows.

We can explain the last difference between the submagnetosonic and
supermagnetosonic cases (compare Figs.~3 and 4) by analogy with a
forced harmonic oscillator, whose response is in phase or out of phase
with the external sinusoidal forcing depending on whether the forcing
frequency is lower or higher than the natural frequency.  A similar
effect evidently distinguishes the ability of fluid elements to respond
in or out of phase to the nonaxisymmetric forcing of the collective
gravitational potential depending on whether the flow occurs at
submagnetosonic or supermagnetosonic speeds relative to the pattern
speed (zero in the present case).  This distinction could be developed
as a powerful diagnostic of physical conditions in flattened cloud
cores and massive protostellar disks, if both turn out to have lopsided
shapes, because the former can generally be expected to have
submagnetosonic rotation speeds; the latter, supermagnetosonic speeds.

Figure 6 shows additional examples of entirely submagnetosonic flow
(for $D^2=0.5$) and entirely supermagnetosonic flow (for $D^2=1.5$) for
$M=1$ SIDs, but now in the $(u_\varpi,u_\varphi)$ plane.  Models are
computed with different values of $S_1$ by the numerical method
described in \S \ref{sec_NumMeth}.  For comparison, the corresponding
flow solutions obtained with the linear analysis of \S \ref{sec_lin}
are also shown.  Notice that the forced epicyclic motion by the
nonaxisymmetric gravitational field about the gyrocenter marked with a
cross (corresponding to circular motion of the axisymmetric model with
the same value of $D^2$), approaches the magnetosonic transition
(horizontal dashed line) in both cases along a tangent in the
velocity-velocity plane.  This behavior is peculiar to $M=1$ SIDs, and
constitutes a topic to which we will return after discussing the $M>1$
cases.

Figure 7 shows the locus in the $D^2$--$|S_M|$ plane of sequences of
equilibria with given $M$-fold symmetry\footnote{We use here the term {\it sequence}
to indicate a set of neighboring equilibria subject to constraints (on their
Fourier expansion, in the present case).}, ranging from axisymmetric
models ({\it dashed line}) to the points where the submagnetosonic flow
acquires a magnetosonic transition ({\it circles}).  We remind the
reader that, unlike the $M=1$ case, bifurcation of $M > 1$ sequences
from the axisymmetric state occurs at discrete rather a continuum of
values of $D^2$, given by $D^2 = M/(M+2)$.  Thus, $M=2,3,4, \dots$
sequences always begin submagnetosonically, $D^2 < 1$, at $S_M = 0$,
and terminate with a magnetosonic transition (circles) before the
nonlinearity parameter $S_M$ can acquire very large values.

Figure 8 shows iso-surface-density contours and velocity vectors for
$M=2$ equilibria ranging from the axisymmetric limit ($S_2=0$) to just
before the magnetosonic transition ($S_2=0.229$).  Notice the
transformation from oval distortions at small $S_2$ (e.g., $S_2 = 0.1$)
to dumbells at large $S_2$ (e.g., $S_2 = 0.2$).  The latter shapes
terminate at the magnetosonic transition ($S_2=0.229$), where the
pinched neck of the dumbell develops a cusp and the streamlines are
trying to change from circulation around a single center of attraction
to circulation around what looks increasingly like two centers of
attraction.

Figure 9 shows iso-surface-density contours and streamlines for models
with $M$-fold symmetry, $M=2,3,4$ and 5, near the endpoints of the
sequences shown in Figure 7. Finally, Figure 10 shows the
velocity-velocity plots for the same four models.  The solutions with
$M > 1$ in Figure 10 differ from those with $M=1$ in Figure 4 in that
the magnetosonic transition for $M > 1$ are made via the development of
a cusp in both the iso-surface-density and velocity-velocity plots.  We
noted earlier that the magnetosnic transition is made for $M=1$
configurations with the $u_\varpi-u_\varphi$ locus becoming tangent to
the critical curve.

\subsection{Interpretation as Onset of Shocks}
\label{sub_Shocks}

For gas flow in spiral galaxies, Shu et al.~(1973) identified cusp
formation in the velocity-velocity plane, as the onset of a shockwave
with infinitesimal jumps, and we adopt a similar interpretation here.
For trans-magnetosonic flow beyond the cusp solution (not shown in
Figure 10 but see Shu et al.~1973), a smooth transition from
submagnetosonic speeds to supermagnetosonic speeds is possible as the
gas swings toward its closest approach to the center, but a smooth
deceleration from supermagnetosonic speeds back to submagnetosonic
speeds is not possible as this gas climbs outwards and catches up with
slower moving material ahead of it.  The transition to slower speeds is
made instead via a sudden jump (a shockwave of finite strength).  The
shock jump introduces irreversibility to the flow pattern.  Prior to
the appearance of the shockwave, the flow can equally occur in the
reverse direction as in the forward direction, and the streamlines
close on themselves.  After the appearance of a shockwave, time
reversal is no longer possible, and the streamlines no longer close
(see, e.g., the discussions of Kalnajs 1973 and Roberts \& Shu 1973).
Instead, angular momentum is removed from the gas (via gravitational
torques when the patterns of density and gravitational potential show
phase lags) and transferred outward in the disk, causing individual
streamlines to spiral toward the center and increasing the central
concentration of mass.  The problem then becomes intrinsically
time-dependent and cannot be followed by the steady-flow formulation
given in the present paper.

We are uncertain why the magnetosonic transition in the case $M=1$ is
not made via cusp-formation.  It may be that in this special case,
sufficient gravitational deceleration from supermagnetosonic to
submagnetosonic speeds (rather than via pressure forces) can occur as
to allow a smooth trans-magnetosonic flow to occur in a complete
circuit.  Unfortunately, we are unable to study this unprecedented
behavior by the methods of the present paper because the numerical
errors introduced by the truncated Fourier treatment of Poisson's
equation compromise our ability to judge true convergence in these
difficult circumstances.  In any case, it is hard to believe, even if
smooth trans-magnetosonic solutions could be found for lopsided SIDs,
that such solutions could be stable (in a time-dependent sense) to the
creation of shockwaves by small departures from perfect 1-fold
symmetry.

\subsection{Circulation and Energy}
\label{circen}

It is interesting to ask whether the nonaxisymmetric bifurcation
sequences studied in this paper represent merely adjacent equilibria,
or also possible evolutionary tracks that might be accessed by secular
evolution of a single system.  To help answer this question, it is
useful to compute the variation of four quantities along any sequence.
The first quantity is the ratio $C\equiv {\cal C}/{\cal M}$ of the
circulation $\cal C$ associated with a streamline to the mass $\cal M$
that it encloses.  For scale-free equilibria, the value of $C$ is
independent of the spatial location of the streamline used to perform
the calculation.  The second, third, and fourth quantities are the
ratios $T\equiv {\cal T}/{\cal M}$, $P\equiv{\cal P}/{\cal M}$, and $W
\equiv {\cal W}/{\cal M}$, respectively, of the kinetic energy $\cal
T$, pressure work integral $\cal P$, and gravitational work integral
$\cal W$ contained interior to any streamline, to the enclosed mass
$\cal M$.  The quantity $C$ is interesting because Kelvin's circulation
theorem (e.g., Shu 1992) combined with the equation of continuity
states that $C$ is conserved in any time-dependent evolution of an
ideal barotropic fluid\footnote{Kelvin's circulation theorem is not
generally valid in the presence of magnetic fields. However, as shown
by Shu \& Li (1997, see also \S 2 of this paper), the effects of
magnetic tension and magnetic pressure in a SID are equivalent to a
reduction of the gravitational force and an augmentation of gas
pressure by {\it constant} factors. Moreover, because the magnetic
fields connect by assumption in our problem to a vacuum in the vertical
direction above and below the disk, there is no magnetic braking of the
local spin angular momentum of the material in the disk plane, and
Kelvin's circulation theorem remains valid.}.  The quantities $T$, $P$,
and $W$ are interesting because they must satisfy the following scalar
virial theorem (per unit mass):
\be
2T+\Theta P+\epsilon W = 0.
\label{vt}
\ee

Let $\varpi=\varpi_0(\varphi)$ define a streamline in the plane of the
disk. The condition $\Psi= {\rm const}$ in equation (\ref{defPsi})
gives immediately
\be
\varpi_0(\varphi)\propto e^{W(\varphi)/D},
\label{defw0}
\ee
where the value of the proportionality constant is irrelevant for what
follows [the reader should not confuse the function $W(\varphi)$ with
$W\equiv {\cal W}/{\cal M}$]. The mass and kinetic energy
contained interior to this streamline are
\be
{\cal M}=\int_0^{2\pi}\int_0^{\varpi_0(\varphi)} \Sigma\varpi d\varpi d\varphi,
\ee
\be
{\cal T}={1\over 2}\int_0^{2\pi}\int_0^{\varpi_0(\varphi)}
\Sigma (u_\varpi^2+u_\varphi^2)\varpi d\varpi d\varphi,
\ee
whereas the circulation and pressure and gravitational work
integrals associated with this streamline are
\be
{\cal C}=\oint {\bf u}\cdot d{\bf l}=\int_0^{2\pi}
\left(u_\varpi {d\varpi_0\over d\varphi} + u_\varphi\varpi_0\right) d\varphi,
\ee
\be
{\cal P} = -\int\int {\bf x}\cdot \nabla \Pi\, d^2x =
-a^2 \oint d\varphi\int_0^{\varpi_0(\varphi)} \varpi {\partial \Sigma
\over \partial \varpi} \, \varpi d\varpi ,
\ee
\be
{\cal W} = -\int\int \Sigma {\bf x}\cdot \nabla {\cal V}\,  d^2x
= -\oint d\varphi \int_0^{\varpi_0(\varphi)} \varpi {\partial {\cal V}\over
\partial \varpi} \Sigma \, \varpi d\varpi.
\ee
Notice that the quantity $\cal P$ equals twice the thermal energy
minus a surface term only if we perform an integration by
parts, which we do not do here (cf. \S 3.2 in Paper I).

If we introduce the nondimensional variables defined in \S 2.2, these
expressions become
\be
{\cal M}=KI_1,
\qquad {\cal T}={1\over 2}K\Theta a^2 I_2,
\ee
\be
{\cal C}={\Theta^{1/2}a\over D} I_2,
\qquad {\cal P} = a^2K I_1,
\qquad \qquad {\cal W}=-2\pi G K^2 I_1,
\ee
where we have used equation (\ref{gravpot}) to evaluate
$\varpi \partial {\cal V}/\partial \varpi$ as $2\pi G K$,
and where
\be
I_1\equiv \int_0^{2\pi}S\varpi_0\; d\varphi,\qquad I_2\equiv
\int_0^{2\pi}{U^2+D^2\over S}\varpi_0\; d\varphi.
\ee
Multiplying equation~(\ref{gs}) by $\varpi_0(\varphi)$
defined by equation~(\ref{defw0}) and integrating 
over a complete cycle, we obtain
\be
{I_2\over I_1}=DB.
\ee
Therefore,
\be
C\equiv {{\cal C}\over {\cal M}} 
= {2\pi\epsilon G\over \Theta^{1/2}a}\left( {B\over 1+DB}\right),
\ee
and 
\be
P\equiv {{\cal P}\over {\cal M}}= a^2,
\qquad T\equiv {{\cal T}\over {\cal M}}={\Theta a^2\over 2} DB,
\qquad W \equiv {{\cal W}\over {\cal M}}=-{\Theta a^2\over \epsilon}(1+DB),
\label{ptw}
\ee
where we have used equation (\ref{defK}) to eliminate $K$.  With the
expressions (\ref{ptw}), the scalar virial theorem (\ref{vt}) is satisfied identically.

Since $M=1$ equilibria exist as a densely populated set of points in
the $D^2$--$|S_1|$ plane, it is clear that we can choose many sequences
for them that have constant values for $C$.  For fixed $\lambda$ (field
freezing) and $a$ (isothermal systems), $C$ is constant along curves of
constant $B/(1+DB)$ = $D_0/(1+D_0^2)$, where $D_0$ is the axisymmetric
value of $D$.  
Thus, on such a sequence,
\be
BD = {D_0D\over 1+D_0(D_0-D)}.
\ee

The dotted curves in Figure 3 show such loci for two representative
sequences in the $D^2$--$|S_1|$ plane:  one submagnetosonic, the other
supermagnetosonic.  At the beginning and end of the supermagnetosonic
sequence displayed in Figure 3, $D_0^2 = 1.50$ and $D^2=1.84$.  Hence,
$BD$ varies from 1.50 at the beginning to 1.98 at the end, and $-W
\propto (1+BD)$ therefore increases by about 19\% from beginning to
end.  In other words, rapidly rotating, self-gravitating SIDs with
diplaced centers are more gravitationally bound than their axisymmetric
counterparts.  In the presence of dissipative agents that lower the
energy while preserving the circulation, such disks will secularly tend
toward greater asymmetric elongation (see Fig. 5).  More gravitational
energy is released when distorted streamlines bring matter closer to
the center than is expended when the same streamlines take the matter
farther from the center, conserving circulation.  This exciting result
deserves further exploration both theoretically and observationally for
systems other than the full singular isothermal disk.

At the beginning and end of the submagnetosonic sequence displayed in
Figure 3, $D_0^2=0.60$ and $D^2=0.35$.  Hence, $BD$ varies from 0.60 at
the beginning to 0.40 at the end, and $-W \propto (1+BD)$ therefore
decreases by about 12\% from beginning to end.  This variation is not
very much considering how fast this sequence rotates relative to
realistic cloud cores.  Nevertheless, the formal decrease of $-W$ as
one leaves the axisymmetric state implies that submagnetosonic systems
require some input of energy to make them less round.  Exceptions are
sequences that branch from smaller values of $D_0^2$, which have
smaller variations of $-W$.  In particular, long spindles have no
binding energy disadvantage whatsoever relative to axisymmetric disks
for the nonrotating sequence shown in Figure 1, because here $-W
\propto (1+BD) = 1$, a constant.  In this regard, it may be significant
that observed cores that are significantly lopsided (see Fig. 2)
typically rotate quite slowly.

The story is more ambiguous for $M > 1$ equilibria.  Here, for given
$M$, the stationary states occupy one-dimensional curves in the
$D^2$--$|S_M|$ plane; therefore, we have no control over how $C$ and
$-W$ vary along any sequence.  Plotted in Figure 11 are the values of
$C$ and $-W$ as we vary $S_M$ along the sequences for $M=$ 2, 3, 4, 5.
Amazingly, the normalized circulation $C$ is nearly, but not exactly,
constant on each sequence, varying by no more than 1\% in all cases.
Given the small values of $S_M$ for which solutions exist and the
relatively small variation of $D$ along each sequence, this result is
not surprising, because $B$ and $DB$ differ from their values for
axisymmetric SIDs by terms ${\cal O}(S_M^2)$.   Although in principle
secular evolution along any $M>1$ sequence would require a slight
redistribution of circulation with mass, the amount required is truly
slight, and one could imagine that mechanisms might exist that can
effect a slow transformation along the sequence toward more
nonaxisymmetric states.  In principle, such evolution would seem to
favor the formation of $M$ = 2, 3, 4, 5, \ldots buds, depending on the
rate of rotation present in the underlying flow.  However, before 2, 3,
4, 5, \ldots independently orbiting bodies can form by such a
``fission'' process, this sequence of events would terminate in
shockwaves, and the resultant transfer of angular momentum (or
circulation) outward and mass inward would stabilize the system against
actual successful fission.

The non-constancy of $C$ along sequences of equilibria with fixed $M$
is reminescent of the change of equatorial circulation along the Jacobi
sequence of incompressible ellipsoids at fixed $J$ and $M$
(Lynden-Bell~1965). A frictionless body is incapable of evolving along
the Jacobi sequence, and a Maclaurin spheroid cannot fall into a Jacobi
ellipsoid, in agreement with the known stability of the frictionless
spheroid at the bifurcation point \footnote{We thank the referee for
suggesting this analogy. Theorems on the stability of sequences of
equilibria defined by a given value of $C$ were derived by Lynden-Bell
\& Katz~(1981).}. In practice, for gaseous systems, a more practical
difficulty mitigates against even beginning the secular paths of
evolution described in the previous paragraphs for the submagnetosonic
cases.  The nonaxisymmetric SIDs with $M=2,3,4,5$ depicted in Figures 8
and 9 are all rotating too slowly to be stable against ``inside-out''
collapse of the type studied for their axisymmetric counterpart by Li
\& Shu (1997).  This dynamical instability would formally overwhelm any
secular evolution along the lines described above.  (Supermagnetosonic
$M=1$ configurations rotate quickly enough to be stable against
``inside-out'' dynamical collapse, and a secular transformation to the
more elongated and eccentrically displaced states of Fig. 5 are
realistic theoretical possibilities.) We plan to study the dynamical
collapse and fragmentation properties of nonaxisymmetric,
submagnetosonic SIDs with general $M$-fold symmetry in a future paper.
In another treatment, we shall also discuss the question whether
configurations with strict $M > 1$ symmetry are formally (secularly)
unstable also to perturbations of $M=1$ periodicity (i.e., to
additional ``lopsided'' bifurcations).  But, for the present, we merely
remark that the practical attainment of any of the nonaxisymmetric
pivotal states depicted, say, in Figure 8 probably occurs, not along a
sequence where each member has already achieved a (nearly) singular
value of surface density at the origin $\varpi = 0$, but along a line
of evolution (perhaps by ambipolar diffusion) where the growing central
concentration of matter occurs without the a priori assumption of axial
symmetry (e.g., nonaxisymmetric generalizations of the calculations of
Basu \& Mouschovias 1994).

\section{Summary and Discusssion}
\label{sumdisc}

In this paper we have shown that prestellar molecular cloud cores
modeled in their pivotal state just before the onset of gravitational
collapse (protostar formation and envelope infall) as magnetized
singular isothermal disks need not be axisymmetric.  The most
impressive distortions are those that make slowly rotating circular
cloud cores lopsided ($M=1$ asymmetry).  Although slowly rotating,
lopsided cloud cores have a slight disadvantage relative to their
axisymmetric counterparts from an energetic point of view, such
elliptical configurations do seem to appear in nature (see Fig. 2).

More intriguingly, elongated, eccentrically displaced,
supermagnetosonically rotating SIDs (that are stable to overall
graviational collapse) are preferred to their axisymmetric counterparts
if the excess binding energy of the latter can be radiated away without
changing the circulation of the streamlines.  If the mass of the
circumstellar disk of a very young protostar is a large fraction of the
mass of the system, it might be possible to find such $M=1$ distortions
of actual objects by future ALMA observations.  If such disks have
(perturbed) flat rotation curves, we predict (see Fig. 5) that the
mm-wave isotphotes should be elongated perpendicular to an
eccentrically displaced central star and also perpendicular to the
eccentric shape of the streamlines (as might be deducible from
isovelocity plots common for investigations of spiral and barred
galaxies).

Bifurcations into sequences with $M =2$, 3, 4, 5, and higher symmetry
require rotation rates considerably larger ($> 0.7$ times the
magnetosonic speeds) than is typically measured for observed molecular
cloud cores (e.g., Goodman et al.~1993).  Although seemingly more
promising for binary and multiple star-formation, the models with
$M=2$, 3, 4, 5, ... symmetries all terminate in shockwaves before their
separate lobes can succeed in forming anything that resembles separate
bodies (see Fig.~8).  For these configurations to exist at all, the
basic rotation rate has to be fairly close to magnetosonic.  It is then
not possible for the nonaxial symmetry to become sufficiently
pronounced as to turn streamlines that circulate around a single center
to streamlines that circulate around multiple centers (as is needed to
form multiple stars), without the distortions causing
supermagnetosonically flowing gas to slam into submagnetosonically
flowing gas.  The resultant shockwaves then increase the central
concentration in such a fashion as to suppress the tendency toward
fission.

We have managed to gain the above understanding semi-analytically only
because of the mathematical simplicity of isopedically magnetized
SIDs.  The same understanding probably underlies similar findings from
numerical simulations of the fission process that inevitably end with
the creation of shockwaves before the actual production of two or more
separately gravitating bodies (Tohline 2000, personal communication).
This negative result, combined with the analysis of the spiral
instabilities that afflict the more rapidly rotating, self-gravitating,
disks into which more slowly rotating, cloud cores collapse (also
modeled here as SIDs), is cause for pessimism that a successful
mechanism of binary and multiple star-formation can be found by either
the fission or the fragmentation process acting in the aftermath of the
gravitational collapse of marginally supercritical clouds {\it during
the stages when field freezing provides a good dynamical assumption}.

It might be argued that our analysis also assumed smooth starting
conditions, and that therefore, turbulence might be the more important
missing ingredient.  However, the low-mass cloud cores in the Taurus
molecular cloud that gives rise to many binaries and multiple-star
systems composed of sunlike stars are notoriously quiet, with turbulent
velocities that are only a fraction of the thermal sound speed (e.g.,
Fuller \& Myers 1992).  Such levels of turbulence are well below those
that appear necessary to induce ``turbulent fragmentation'' in the
numerical simulations of Klein et al.~(2000).  Interstellar turbulence
is undoubtedly an important process at the larger scales that
characterize the fractal structures of giant molecular clouds (see,
e.g., Allen \& Shu 2000), but it probably plays only a relatively minor
role in the simplest case of isolated or distributed star-formation
that we see in clouds like those in the Taurus-Auriga region, which
has, as we mentioned earlier, more than its share of cosmic binaries.

In contrast, we know that the dimensionless mass-to-flux ratio
$\lambda$ had to increase from values typically $\sim 2$ in cloud cores
to values in excess of 5000 in formed stars (Li \& Shu 1997).  Massive
loss of magnetic flux must have occurred at some stage of the
gravitational collapse of molecular cloud cores to form stars.
Moreover, this loss must take place at some point at a dynamical rate,
or even faster, since the collapse process from pivotal molecular cloud
cores is itself dynamical.  It is believed that dynamical loss of
magnetic fields from cosmic gases occurs only when the volume density
exceeds $\sim 10^{11}$ H$_2$ molecules cm$^{-3}$ (e.g., Nakano \&
Umebayashi 1986a,b; Desch \& Mouschovias 2000).  It might be thought
that cloud cores have to collapse to fairly small linear dimensions
before the volume density reaches such high values, and therefore, that
only close binaries can be explained by such a process, but not wide
binaries (McKee 2000, personal communication).  However, this
impression is gained by experience with {\it axisymmetric} collapse.
Once the restrictive assumption of perfect axial symmetry is removed,
we gain the possibility that some dimensions may shrink faster than
others (e.g., Lin, Mestel, \& Shu 1965), and densities as high as
$10^{11}$ cm$^{-3}$ might be reached while only one or two dimensions
are relatively small, and while the third is still large enough to
accomodate the (generally eccentric) orbits of wide binaries.

We close with the following analogies.  The basic problem with trapped
magnetic fields is that they compress like relativistic gases (i.e.,
their stresses accumulate as the 4/3 power increase of the density in
3-D compression).  Such gases have critical masses [e.g., the
Chandrasekhar limit in the theory of white dwarfs, or the magnetic
critical mass of equation~(\ref{MPhi})] which prevent their
self-gravitating collections from suffering indefinite compression, no
matter how high is the surface pressure, if the object masses lie below
the critical values.  Moreover, while marginally supercritical objects
might collapse to more compact objects (e.g., white dwarfs into neutron
stars, or cloud cores into stars), a single such object cannot be
expected to naturally fragment into multiple bodies (e.g., a single
white dwarf with mass slightly bigger than the Chandrasekhar limit into
a pair of neutron stars).

In order for fragmentation to occur, it might be necessary for the
fluid to decouple rapidly from its source of relativistic stress.  For
example, the universe as a whole always has many thermal Jeans masses.
Yet in conventional big-bang theory, this attribute did not do the
universe any good in the problem of making gravitationally bound
subunits, as long as the universe was tightly coupled to a relativistic
(photon) field.  Only after the matter field had decoupled from the
radiation field in the recombination era, did the many fluctuations
above the Jeans scale have a chance to produce gravitational
``fragments.'' It is our contention that this second analogy points
toward where one should search for a viable theory of the origin of
binary and multiple stars from the gravitational collapse of magnetized
molecular cloud cores.

\acknowledgements

We thank an anonymous referee for useful comments on the manuscript. We
also thank the referee of Paper I for suggesting that we examine the
variation of $C$ in our bifurcation sequences as a discriminant between
equilibria that merely lie adjacent to each other in parameter space
and states that can be connected by a secular line of evolution.  We
also wish to express our appreciation to Tim de Zeeuw, Chris McKee,
Steve Shore, Joel Tohline for insightful comments and discussions, and
to Fr\'ed\'erique Motte for providing the map of L1544 at 1.3~mm.  The
research of DG is partly supported by ASI grant  ARS-98-116 to the
Osservatorio di Arcetri. The research of FHS and GL is funded in part
by a grant from the National Science Foundation and in part by the NASA
Astrophysical Theory Program that supports a joint Center for Star
Formation Studies at NASA Ames Research Center, the University of
California at Berkeley, and the University of California at Santa
Cruz.  SL acknowledges support from DGAPA/UNAM and CONACyT.

\appendix
\section{Appendix}
\label{appen}

Given the definition 
\be
L_m \equiv {1\over 2\pi}\oint \ln (1-\cos \varphi)
\cos(m\varphi) \, d\varphi= 
{1\over \pi}\int_0^\pi \ln (1-\cos \varphi)
\cos(m\varphi) \, d\varphi, 
\label{eq0}
\ee
where $m$ is an integer (positive or negative), we show that
\be
L_m = \left\{
\begin{array}{ll}
-\ln 2 & \mbox{if $m=0$},       \\
-1/|m| & \mbox{if $|m| \ge 1$}.
\end{array}
\right.
\ee

(1) For $m=0$, 
\be
L_0 = {1\over 2\pi} \oint \ln(1-\cos \varphi)\, d\varphi = 
{1\over \pi}\int_0^\pi \ln(1-\cos \varphi) \, d\varphi 
\label{eq1}
\ee
Successive transformations $\varphi
\rightarrow -\varphi$ and $\varphi \rightarrow \varphi+\pi$ demonstrate
\be
L_0 = {1\over \pi}\int_{-\pi}^0 \ln(1-\cos\varphi)\, d\varphi
={1\over \pi}\int_0^\pi \ln (1+\cos\varphi) \, d\varphi,
\label{eq2}
\ee
which is just a statement that $\cos \varphi$ is odd
relative to the midpoint of the interval $(0,\pi)$.
If we add equations (\ref{eq1}) and (\ref{eq2}), we get
\be
2L_0 = {1\over \pi}\int_0^\pi \ln(1-\cos^2\varphi) \,
d\varphi = {1\over \pi}\int_0^\pi \ln(\sin^2 \varphi) \,
d\varphi.
\ee
Rewrite $\sin^2 \varphi = (1/2)(1-\cos \chi)$ where
$\chi \equiv 2\varphi$; then
\be
2L_0 = -\ln 2 +{1\over 2\pi}\int_0^{2\pi} \ln(1-\cos\chi) \, d\chi .
\ee
But the last integral is another expression for $L_0$
(see eq.~[\ref{eq1}]); thus, $L_0 = -\ln 2$.  (QED)

(2) Integrate equation (\ref{eq0}) once by parts to obtain
\be
mL_m = -{1\over 2\pi}\int_{-\pi}^{+\pi} {\sin (m\varphi)
\sin\varphi\, d\varphi \over 1-\cos\varphi} , 
\ee
Multiply and divide the integrand by $1+\cos\varphi$ and write
$1-\cos^2\varphi$ as $\sin^2\varphi$.  Thereby obtain
\be
mL_m = -(I_m + J_m) , 
\label{eq3}
\ee
where
\be
I_m \equiv {1\over 2\pi}\int_{-\pi}^{+\pi} 
{\sin (m \varphi) \, d\varphi \over \sin \varphi} ,
\label{eq4}
\ee
\be
J_m \equiv {1\over 2\pi}\int_{-\pi}^{+\pi}
{\sin (m \varphi) \cos \varphi \, d\varphi \over
\sin \varphi} . 
\label{eq5}
\ee
We easily find
\be
I_1 = 1 \qquad {\rm and} \qquad J_1 = 0 .
\ee

For $m > 1$, write
\be
\sin (m \varphi) = \sin \varphi\cos[(m-1)\varphi] +
\cos \varphi \sin [(m-1)\varphi] .
\ee
Equation (\ref{eq4}) then yields
\be
I_m = J_{m-1} ,
\ee
whereas equation (\ref{eq5}) becomes
\be
J_m = {1\over 2\pi}\int_{-\pi}^{+\pi} \left[
\cos [(m-1)\varphi]\cos\varphi + {\cos^2\varphi
\sin[(m-1)\varphi] \over \sin \varphi}\right] \, d\varphi .
\ee
Write $\cos^2\varphi$ in the second term as $1-\sin^2\varphi$.   Note that
the $-\sin^2\varphi$ term combines with the other term in the integrand
to form $\cos (m\varphi)$, which integrates to zero for $m > 1$.
Thus, obtain
\be
J_m ={1\over 2\pi}\int_{-\pi}^{+\pi} {\sin [(m-1)\varphi]
\, d\varphi \over \sin \varphi} = I_{m-1}.
\ee
Collecting results, we get
\be
I_m + J_m = I_{m-1}+J_{m-1} = \dots =
I_1+J_1 = 1 \qquad {\rm for} \qquad m \ge 1.
\ee
Since $L_m$ is an even function of $m$, equation (\ref{eq3}) now 
implies $L_m=-1/|m|$ for $|m|\ge 1$. (QED)

\clearpage

\begin{figure}
\plotone{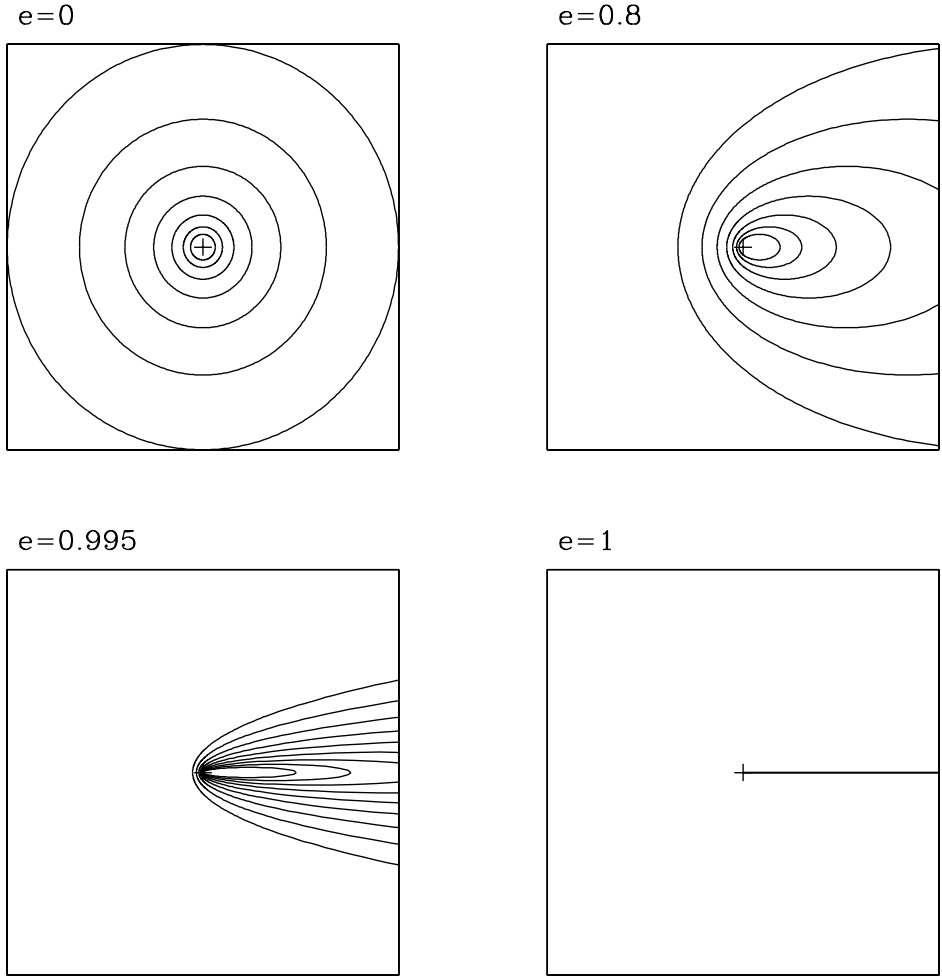}
\caption{Iso-surface-density contours for static SIDs are perfect ellipses
with eccentricity $e$. For $e=0$ the SID is an axisymmetric disk with
surface density $\propto \varpi^{-1}$, for $e=1$ the SID degenerates in
a semi-infinite filament of constant linear mass density.}
\end{figure}
\clearpage

\begin{figure}
\plotone{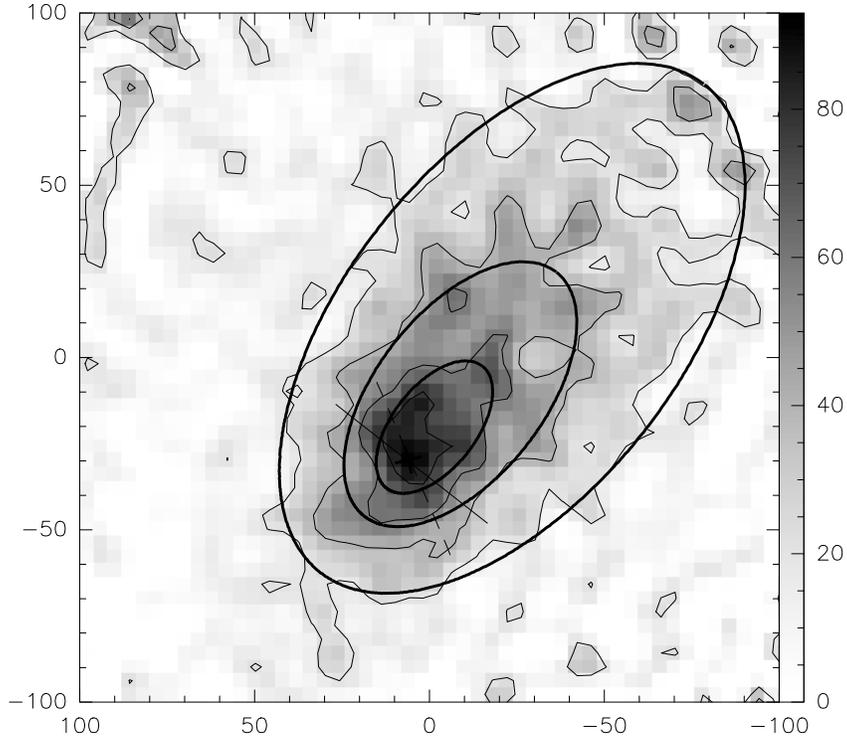}
\caption{Iso-surface-brightness contours ({\it thick solid lines}\/)
from a theoretically computed, lopsided, magnetized, self-gravitating
figure of equilibrium compared with isophotal measurements of
Ward-Thompson et al.~(1999) of the submillimeter emission from heated
dust grains in L1544.  The short {\it solid line} and {\it dashed line}
show the directions of predicted and measured field inferred from
submillimeter-wave polarization observations (Ward-Thompson et
al.~2000).}
\end{figure}
\clearpage

\begin{figure}
\plotone{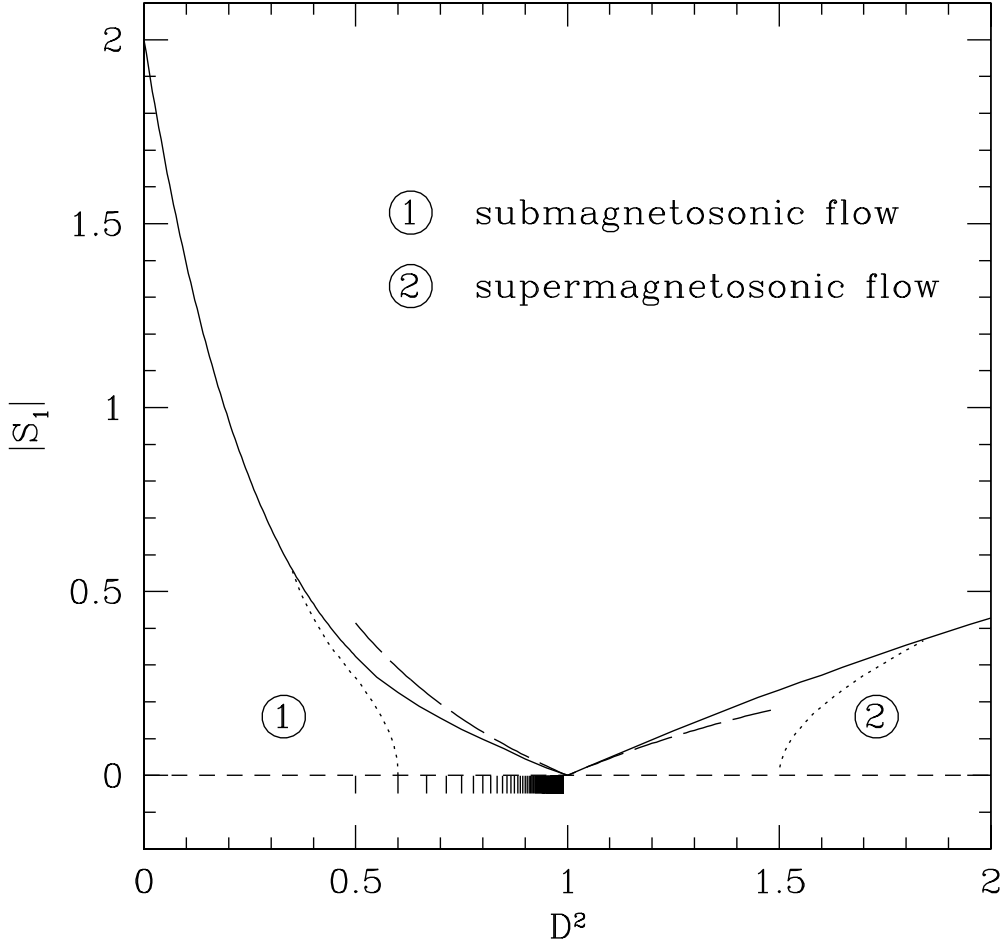}
\caption{The loci (solid curves) in the $(D^2,|S_1|)$ plane of critical
flow solutions for $M=1$ approaching magnetosonic speed starting from
entirely submagnetosonic ($D<1$) or entirely supermagnetosonic flow
($D>1$).  The {\it horizontal short-dashed line} shows the locus of
axisymmetric models.  The {\it long-dashed curves} shows the limit of
magnetosonic models as given by equation (\ref{mcrit}) in the linear
approximation $S_1\ll 1$.  The {\it dotted curves} show a
submagnetosonic and a supermagnetosonic sequence, each of which
maintains a constant ratio $C$ of circulation $\cal C$ to enclosed mass
$\cal M$ (see \S \ref{circen}).  Tickmarks pointing downward from the
horizontal dashed line denote the values of $D^2$ where distortions
with $M$-fold symmetry can occur, with $M > 1$, as predicted by the
linear analysis of Paper I and \S \ref{sec_lin}.}
\end{figure}
\clearpage

\begin{figure}
\plotone{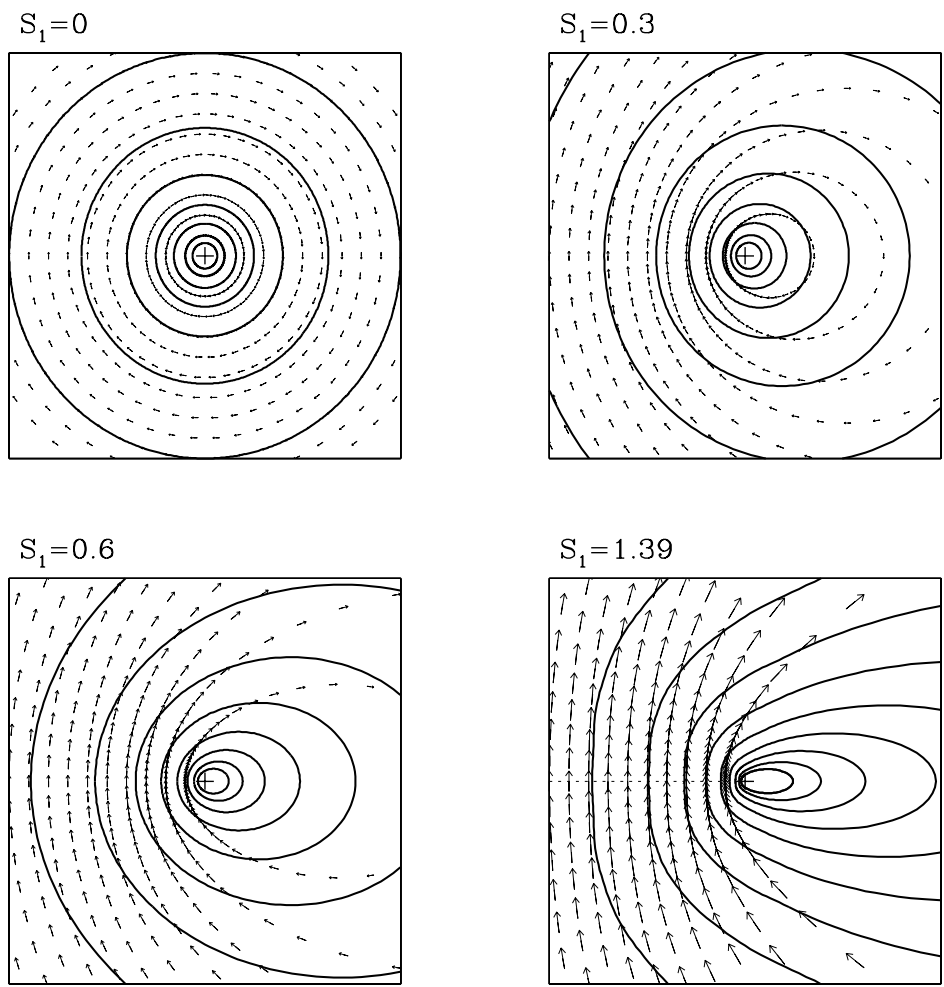}
\caption{Sequence of equilibria with $M=1$ and $D^2=0.1$ for $S_1=0$
(axisymmetric case), 0.3, 0.6 and 1.39 (magnetosonic flow). {\it Thick
lines} correspond to iso-surface-density contours logarithmically
spaced.  Streamlines are outlined by vectors of length proportional to
the modulus of velocity, drawn to scale in the four panels.  Equation
(\ref{ucomp}) shows that the flow velocities depend, in magnitude and
direction, only on the azimuthal angle $\varphi$ and not on the
distance $\varpi$ (for given $\varphi$) from the center of the mass
distribution.  In particular, the gas velocities at perisys (or aposys)
are all the same independent of the size of the streamline.  In the
last panel, the {\it thin dotted line} indicates the locus of
magnetosonic velocity reached at perisys.}
\end{figure}
\clearpage

\begin{figure}
\plotone{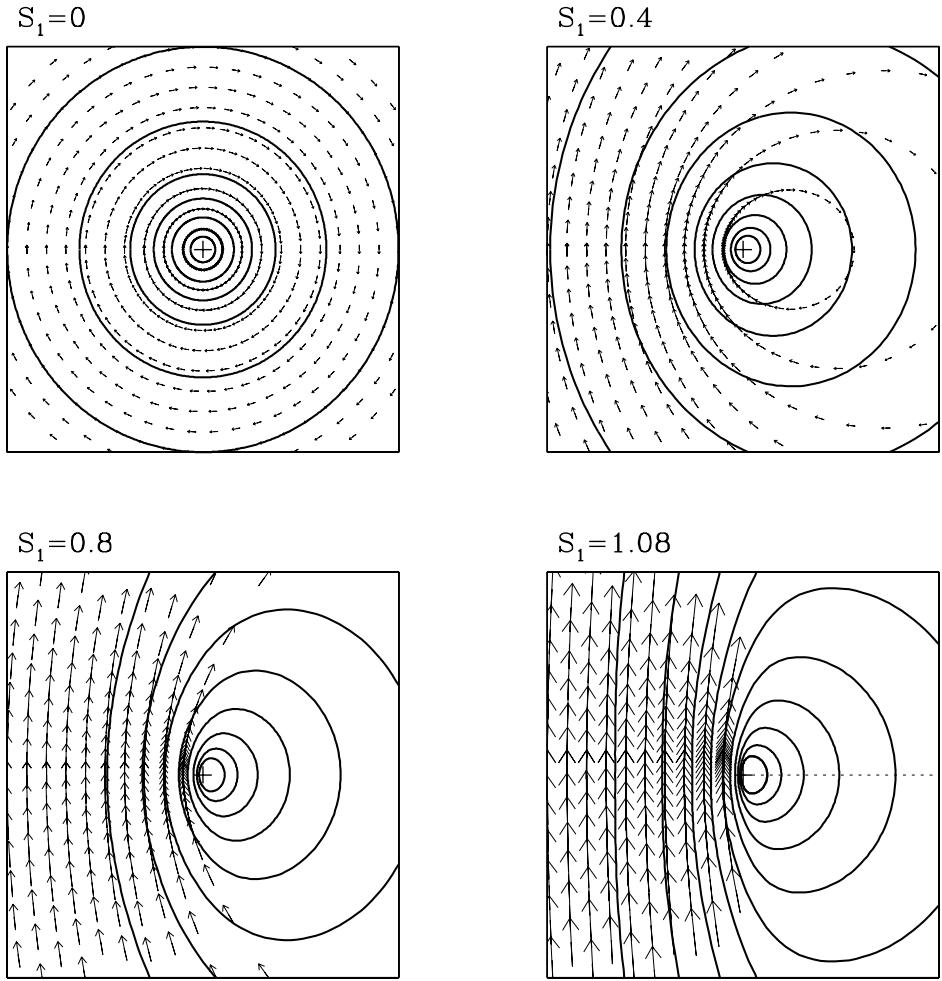}
\caption{Sequence of equilibria with $M=1$ and $D^2=4$ for $S_1=0$
(axisymmetric case), 0.4, 0.8 and 1.08 (magnetosonic flow). Vectors and 
lines are as in Fig.~4.}
\end{figure}
\clearpage

\begin{figure}
\plotone{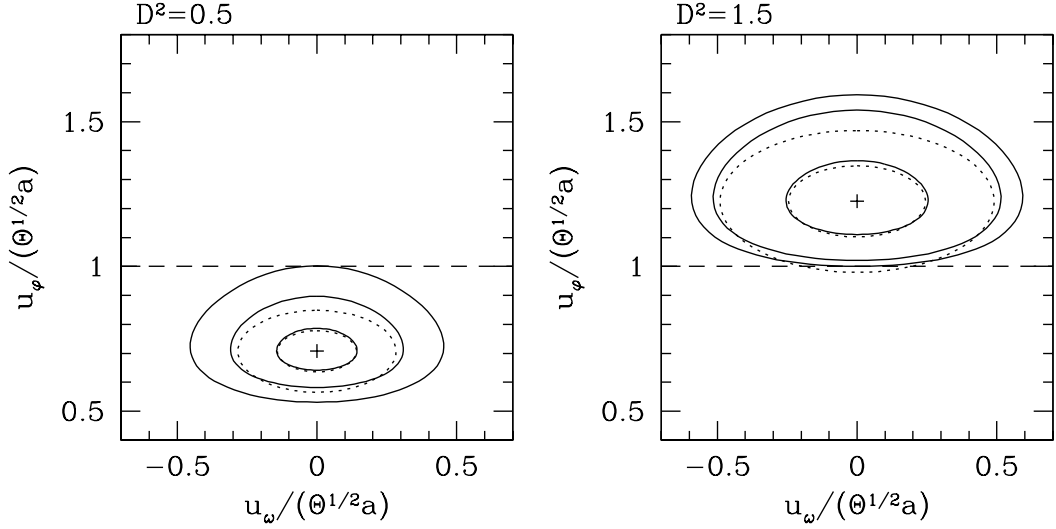}
\caption{Two examples of entirely submagnetosonic flow ($D^2=0.5$) and
entirely supermagnetosonic flow ($D^2=1.5$) in the velocity-velocity
plane, for $M=1$.  In each panel, the axisymmetric solutions is marked
by a cross.  The two inner {\it solid curves} are obtained with
$S_1=0.1$ and $0.2$, whereas the {\it dotted curves} show the
corresponding linearized solutions.  The outermost {\it solid curve},
obtained with $S_1=0.3235$ for $D^2=0.5$ and $S_1=0.2326$ for
$D^2=1.5$, shows the approach to azymuthal magnetosonic flow ({\it dashed line})
defined by the condition $u_\varphi= \Theta^{1/2} a$.}
\end{figure}
\clearpage

\begin{figure}
\plotone{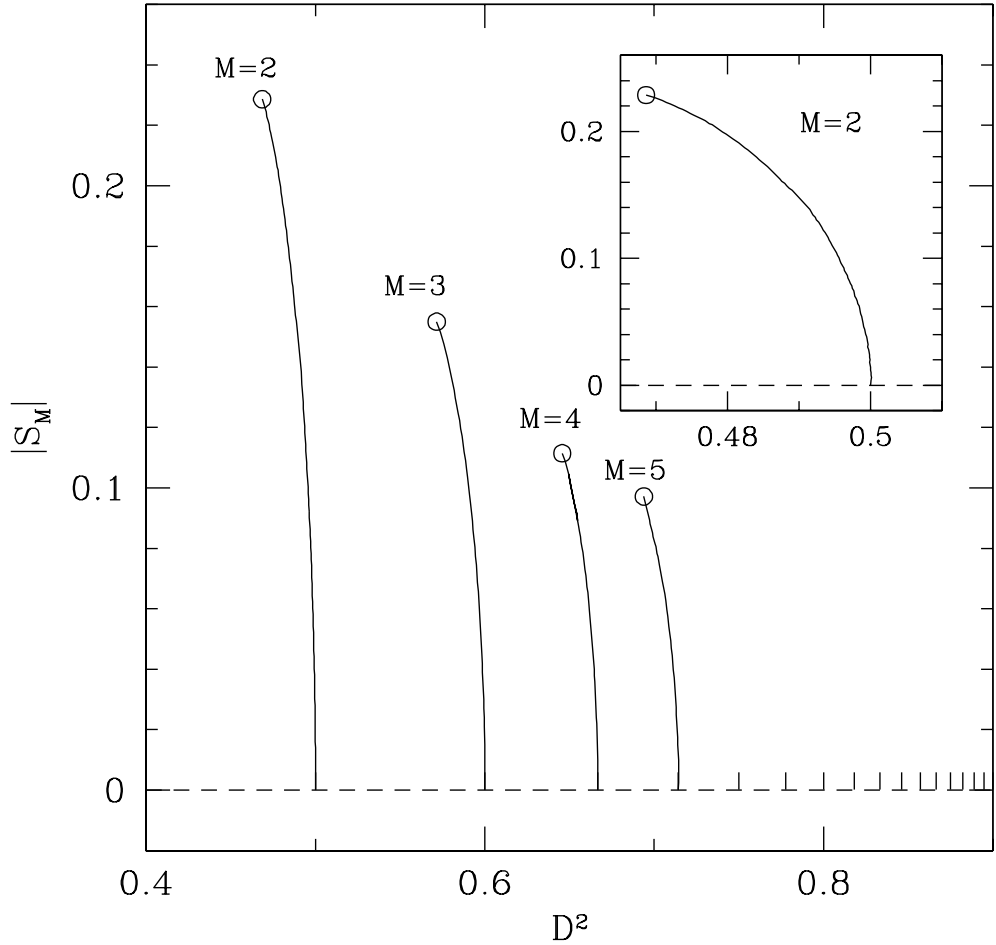}
\caption{Locus in the $D^2$--$|S_M|$ plane of sequences of equilibria
with given $M$-fold symmetry. The {\it dashed line} indicates the locus
of axisymmetric equilibria.  Tickmarks denote the values of $D^2$ where
distortions with $M$-fold symmetry can occur, as predicted by the
linear analysis of Paper I and \S \ref{sec_lin}.  Circles indicate the points
where the sequences terminate because of the occurrence of shocks.}
\end{figure}
\clearpage

\begin{figure}
\plotone{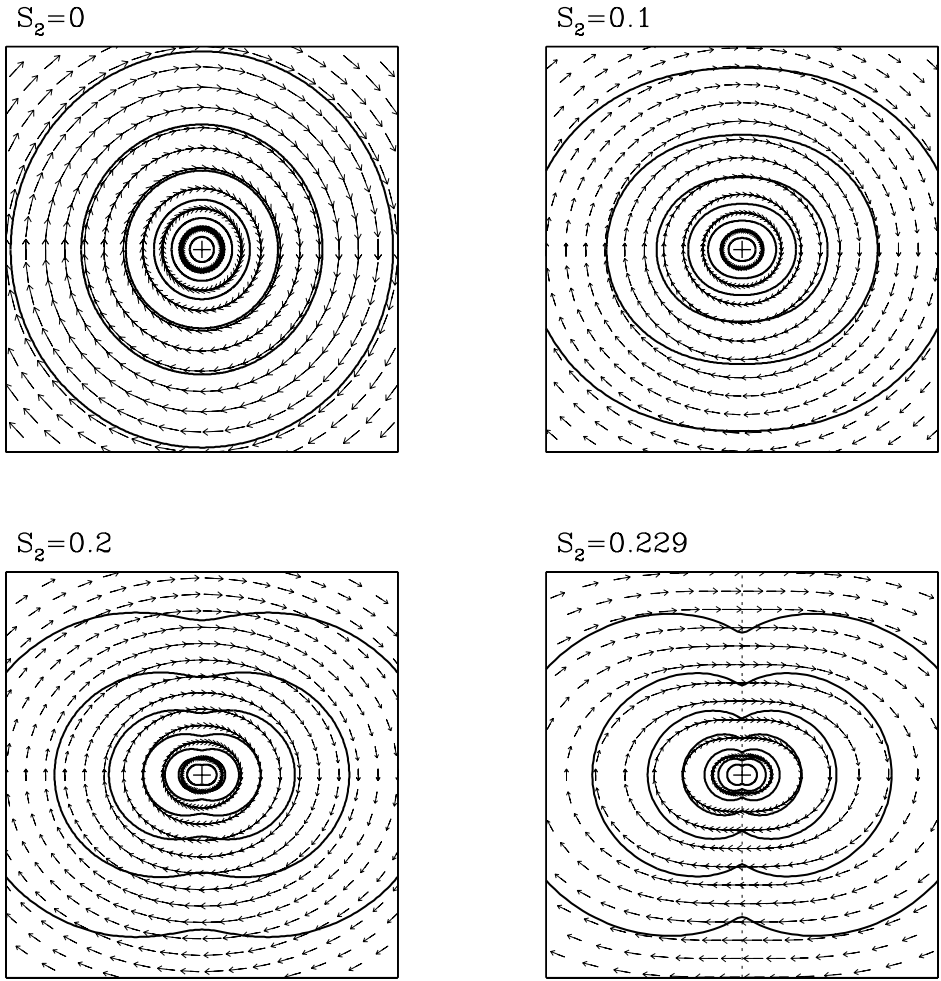}
\caption{Sequence of equilibria with $M=2$, for $S_2=0$ (axisymmetric 
case), 0.1, 0.2 and 0.229 (magnetosonic flow). Vectors and lines are
as in Fig.~4.}
\end{figure}
\clearpage

\begin{figure}
\plotone{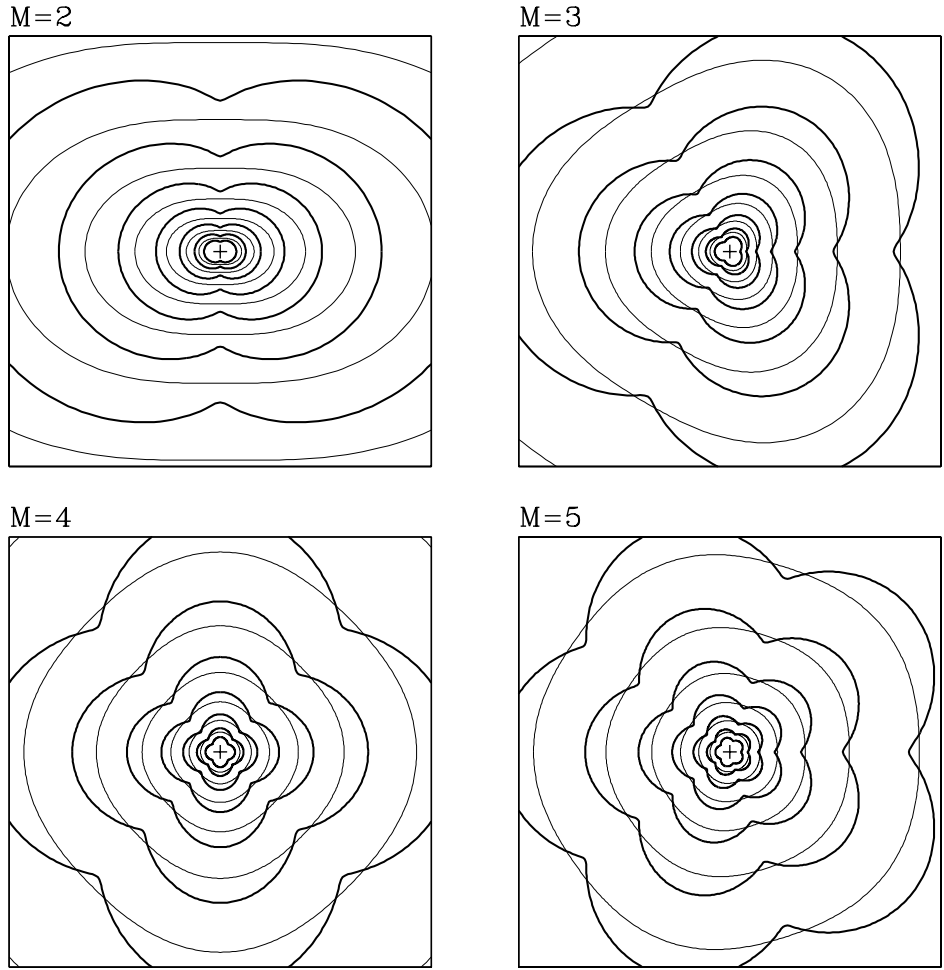}
\caption{Iso-surface-density contours ({\it thick lines}) and streamlines ({\it
thin lines}) for models with $M$-fold symmetry $M=2,3,4$ and 5 at the
point of shock formation.  Iso-surface-density contours develop cusps where the
azymuthal flow velocity reaches the magnetosonic value. }
\end{figure}
\clearpage

\begin{figure}
\plotone{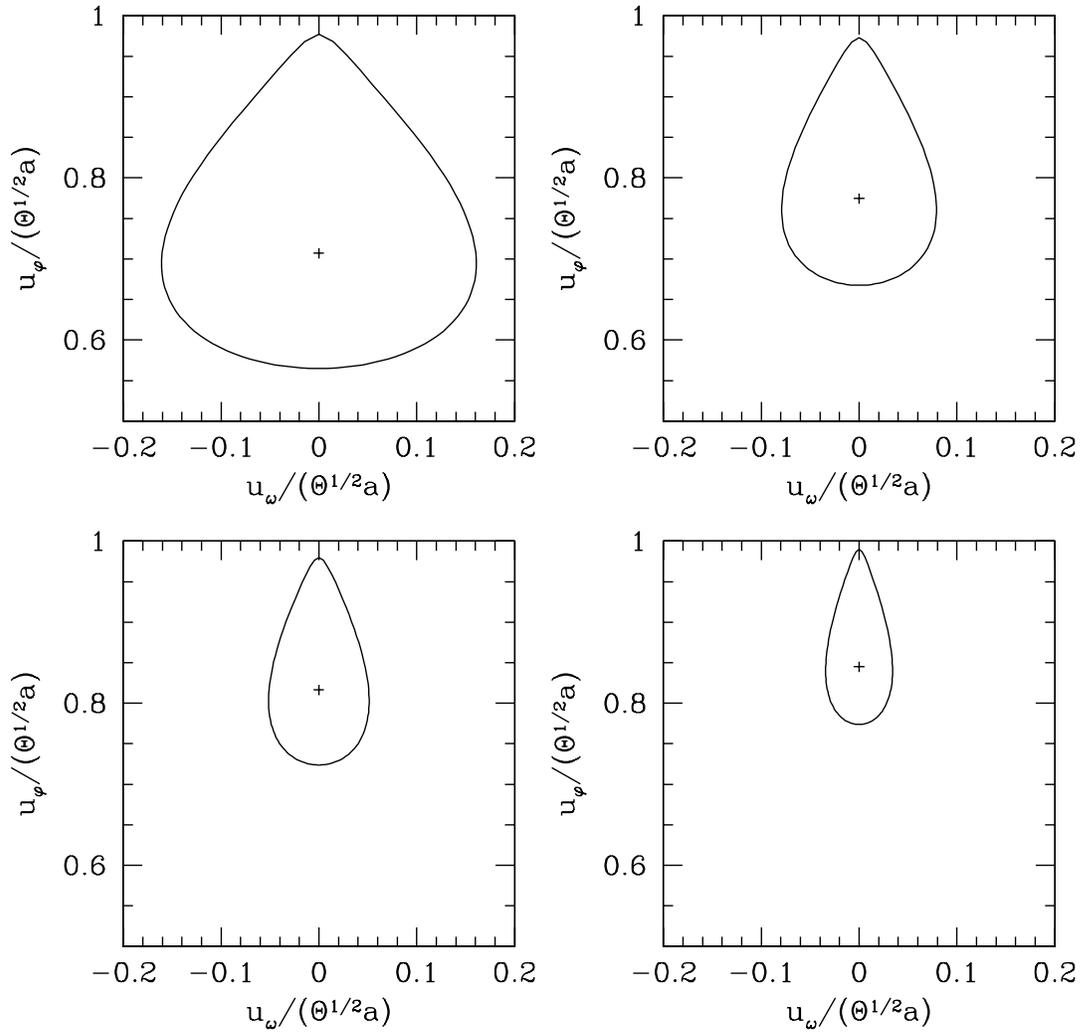}
\caption{Velocity-velocity plots for the four models shown in Fig.~8.}
\end{figure}

\begin{figure}
\plotone{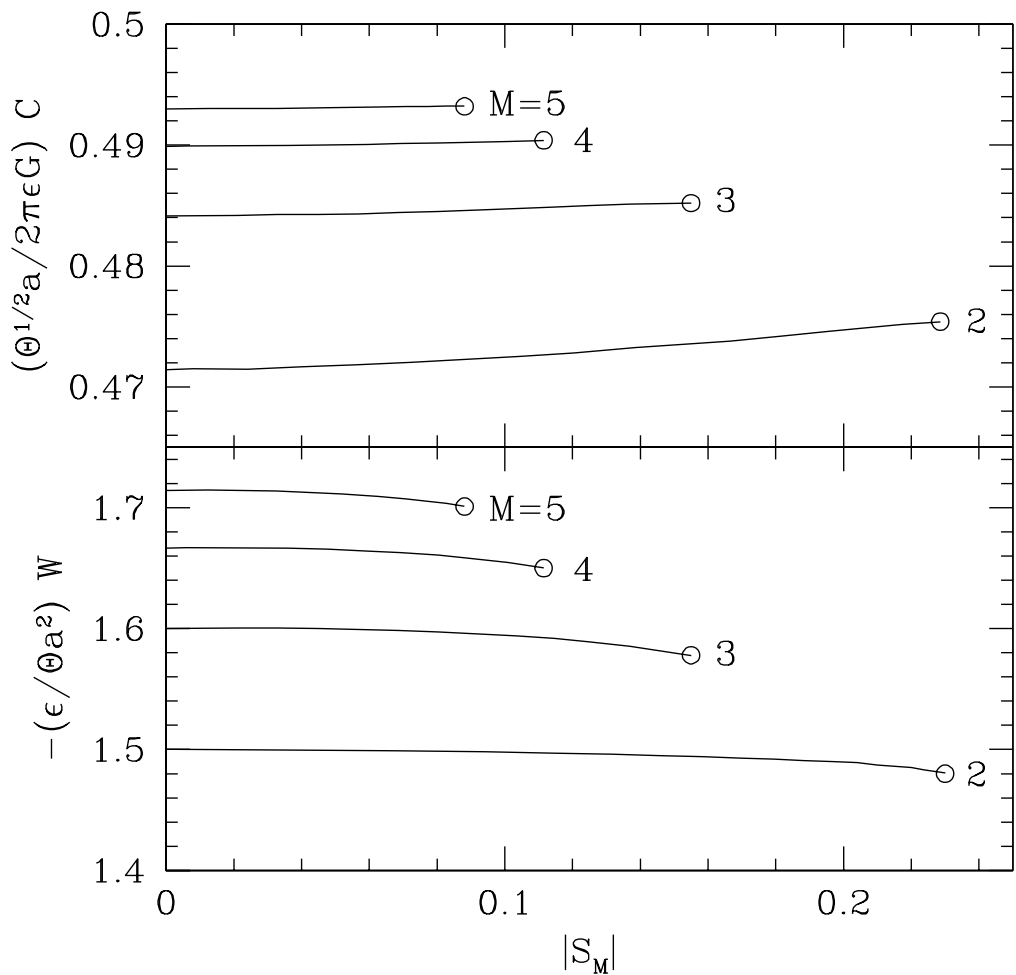}
\caption{Values of $C$ and and $-W$, in units of $2\pi\epsilon
G/\Theta^{1/2}a$ and $\Theta a^2/\epsilon$, respectively, as functions
of $|S_M|$ along the sequences for $M=2$, 3, 4 and 5 shown in
Figure~7. Circles indicate the points where the sequences terminate
because of the occurrence of shocks.}
\end{figure}

\end{document}